\documentclass[aps,prd,twocolumn,superscriptaddress]{revtex4-2}

\usepackage[english]{babel}
\usepackage{amsmath, amsfonts, amsthm, amsbsy, amsopn, amsxtra, amscd, amstext}
\usepackage{longtable}
\usepackage{bbm}
\usepackage{enumitem}
\usepackage{graphicx}
\usepackage{xcolor}
\usepackage{wrapfig}
\usepackage{soul}
\usepackage{physics}
\usepackage{amssymb}

\def\aj{Astronomical Journal}                 
\def\apj{Astrophysical Journal}                
\def\apjl{Astrophysical Journal}

\def\mnras{MNRAS}            
\def\prd{Phys.~Rev.~D}       
\def\prl{Phys.~Rev.~Lett}    
\def\cqg{Class.~Quant.~Grav.~}
\def\araa{ARA\&A}             
\def\nat{Nature}              
\def\aap{A\&A}

\newcommand\E[1]{{\left\langle #1 \right \rangle}}

\newcommand*{\mdeta}{M_{det, A}}
\newcommand*{\mdetb}{M_{det, B}}
\newcommand*{\msrca}{M_{src, A}}
\newcommand*{\msrcb}{M_{src, B}}
\newcommand*{\startrack}{{\tt StarTrack}}
\newcommand*{\msun}{M_{\odot}}
\newcommand*{\bbh}{BBH}
\newcommand*{\bns}{BNS}
\newcommand*{\nsbh}{NSBH}
\newcommand*{\Kbest}{K0559}

\newcommand*{\STFa}{f_a} \newcommand*{\STBeta}{\beta} \newcommand*{\STkick}{\sigma_{\mathrm{eff}}} \newcommand*{\STWindOne}{f_{\mathrm{wind1}}} \newcommand*{\STWindTwo}{f_{\mathrm{wind2}}} \newcommand*{\BinaryParameters}{\vec{\lambda}}
\newcommand*{\AllBinaryParameters}{\{\vec{\lambda}\}}
\newcommand*{\FormationParameters}{\Lambda}

\newcommand*{\pprob}{\mathrm{P}}

\newcommand*{\redshift}{z}
\newcommand*{\metallicity}{\mathcal{Z}}
\newcommand*{\gwdetj}{d_j}
\newcommand*{\gwdets}{\{\gwdetj\}}
\newcommand*{\lumdist}{l}

\newcommand*{\kickAdj}{substantial}
\newcommand*{\kick}{SN recoil kick}
\newcommand*{\kicks}{SN recoil kicks}
\newcommand*{\engine}{delayed}
\newcommand*{\snr}{\mathrm{SNR}}
\newcommand*{\lalsnr}{\mathrm{SNR}_{\mathrm{lal}}}

\newcommand\mc{{\cal M}_c}
\newcommand\HideMe[1]{}
\newcommand\unit[1]{{\rm #1}}

\newcommand\NsimFull{442{} }

\newcommand\citeRatesField{\cite{LIGO-Inspiral-Rates,PSgrbs-popsyn,Belczynski2010metallicity,DominikI,DominikII}}

\newcommand\citeBNS{\cite{abbott2017gw170817}}
\newcommand\citeNSBH{\cite{LIGO-O3-O3b-NSBH}}
\newcommand\citeOSG{\cite{osg07,osg09}}
\newcommand\citeMCMCpopsyn{\cite{Talbot2019GWPopulation, Wysocki2018, Wysocki2019, LIGO-O3-O3b-RP,
    Belczynski2020-EvolutionaryRoads,Breivik_2020, Sadiq2021, Edelman2021, Tiwari_2021}}
\newcommand\citeKDEpopsyn{\cite{Ghosh_2021}}
\newcommand\citeGMMpopsyn{\cite{Golomb_2022}}

\newcommand\citeIsolated{\cite{1991ApJNarayan, Belczynski2002, Belczynski2008, Belczynski2016, Belczynski2020-EvolutionaryRoads, broekgaarden2021formation, posydon, Stevenson2022, Bulik2003, Bulik2004, COSMICWong2021, COSMICZevin2021, COSMICWong2022}}
\newcommand\citeBE{\cite{Izzard2004, Izzard2006, Izzard2009, Eldridge2017, Vanbeveren1998a, Vanbeveren1998b, Hurley2002, Kruckow2018, Breivik_2020, Giacobbo2017, Lipunov2009, Spera2015, Toonen2012, Toonen2016}}

\newcommand\citeWeakWind{\cite{Bouret2005, Bouret2012, Surlan2013}}
\newcommand\citeWeakWindSMC{\cite{Ramachandran2019,Bjorklund2021Wind,Yarovova_2022,Rickard_2022}}
\newcommand\citeWeakWindFactor{\cite{GormazMatamala2019, GormazMatamala2022, GormazMatamala2021, Bjorklund2021Wind,Sundqvist2019, Kritcka2017, Gayley2022}}

\newcommand\citeBetterKicks{\cite{Belczynski2020-EvolutionaryRoads, Olejak2022, Fryer2012, MandelKick, Giacobbo2017, Bray2016Kick, VignaGomez2018, Nordhaus2012}}

\newcommand{\AffiliationCCRG}{
  Center for Computational Relativity and Gravitation, 
  Rochester Institute of Technology, 
  Rochester, New York 14623, USA 
}
\newcommand{\AffiliationWarsawU}{
  Astronomical Observatory, 
  Warsaw University, 
  Al. Ujazdowskie 4, 00-478 Warsaw, Poland 
}
\newcommand{\AffiliationPAS}{
  Nicolaus Copernicus Astronomical Center, 
  Polish Academy of Sciences, 
  ul. Bartycka 18, 00-716 Warsaw, Poland 
}
\newcommand{\AffiliationUWM}{
  Department of Physics, 
  University of Wisconsin--Milwaukee, 
  Milwaukee, WI 53201, USA 
}
\newcommand{\AffiliationGSFC}{
  Gravitational Astrophysics Laboratory, 
  NASA Goddard Space Flight Center, 
  Greenbelt, MD 20771, USA 
}

\begin{document}
\title{
       Iteratively Comparing Gravitational-Wave Observations\\
       to the Evolution of Massive Stellar Binaries
}
\author{V. Delfavero}
\email[]{vera.delfavero@nasa.gov}
\affiliation{\AffiliationGSFC}
\affiliation{\AffiliationCCRG}

\author{R. O'Shaughnessy}
\affiliation{\AffiliationCCRG}

\author{K. Belczynski}
\affiliation{\AffiliationPAS}

\author{P. Drozda}
\affiliation{\AffiliationWarsawU}

\author{D. Wysocki}
\affiliation{\AffiliationUWM}

\date{\today}

\clearpage{}\begin{abstract}
Gravitational-wave observations have the capability to strongly differentiate
between different assumptions
    for how binary compact objects form.
The agreement of observations to 
    different models of the evolution of massive stellar binaries
    leading to the formation of compact binaries
    can be characterized by a Bayesian marginal likelihood.
In this work, we show how to carefully interpolate this
    marginal likelihood between choices of binary evolution model parameters,
    enabling the analysis of their posterior distributions
    between expensive binary evolution simulations.
Using the \startrack{} binary evolution code,
    we compare one- and four-dimensional binary evolution models to the compact
    binary mergers reported in recent gravitational-wave observing runs,
    considering merger detection rates and mass distributions.
We demonstrate that 
    the predicted detection rates and mass distribution of simulated binaries
   can be effective in constraining binary evolution formation.
We first consider a one-dimensional model, studying the effect of
    SuperNova (SN) kick velocity 
    (drawn from a Maxwellian with dispersion $\STkick$)
    on the simulated
    population of compact binary mergers, and find support for \kickAdj{} \kicks{}.
We follow this up with a four-dimensional study of $\STkick$,
    mass transfer efficiency ($\STFa$) and 
the efficiency of angular momentum depletion from ejected material
    ($\STBeta$)
    during Roche-lobe accretion, and an observation-driven reduction
    in the mass-loss rate estimated from stellar wind models ($\STWindOne$).
Of those four formation parameters we investigated,
    we find that three of them ($\STkick$, $\STFa$, and $\STWindOne$)
    can be efficiently limited by these observational comparisons.
After initially sampling from a uniform prior in the space of these parameters,
    we refined our sampling by iteratively
estimating a Bayesian likelihood for each simulation,
    fitting that likelihood to a parametric model (a truncated Gaussian)
    in the four-dimensional formation parameter space,
    and sampling directly from that Gaussian
    in order to propose new simulations in a way that is informed
    by previous simulations.
Our maximum likelihood simulation (K0559) has parameters:
    $\STkick = 108.3$ km/s (indicating \kickAdj{} \kicks{}),
    $\STFa = 0.922$ (indicating efficient mass transfer),
    and $\STWindOne = 0.328$ (indicating support for reduced wind-driven mass loss).
Note that our estimates are only valid within one particular model
    of compact binary formation through isolated binary evolution and
    do not yet take into account the impact of other uncertain pieces of
    stellar physics and binary evolution.
\end{abstract}
\clearpage{}

\maketitle

\section{Introduction}
\label{sec:intro}
By identifying and characterizing the 
    properties of systems which become the
    observed gravitational-wave (GW) sources,
    the distribution of 
    those properties in the population of
    coalescing compact binaries
    can now be measured
    empirically \cite{LIGO-O2-RP, LIGO-O3-O3a-RP,LIGO-O3-O3b-RP,Wysocki2019,IAS, 3-OGC, Boyle_2019_SxS, Jani_2016_GA_Tech, BAM2008}. 
From the first observation of gravitational-waves from the binary black hole
    (\bbh{}) system GW150914
    \cite{GW150914-detection}, 
    has
    quickly challenged candidate models
    for how these compact objects form.
The discovery of this system demonstrated that black holes
    heavier than those observed in X-ray binaries exist
    \cite{miller-jones_2014},
    eliminating models which don't produce them.
Similarly, the observation of the binary neutron star (\bns{}) and neutron star-black hole
    (\nsbh{})
    systems such as GW170817 \citeBNS 
    and GW200105 and GW200115 \citeNSBH, 
    further define the population of these objects which must
    be produced from the massive stars which become compact objects.
The discoveries of GW190412 \cite{GW190412} and GW190814 \cite{GW190814}
    (with mass ratios $m_2/m_1 = 0.28^{+0.12}_{-0.06}$ and $0.112^{+0.008}_{-0.009}$
    respectively)
    similarly demonstrated that asymmetric binaries must be produced,
    eliminating models which can't produce a wide mass ratio spectrum
    \cite{LIGO-O3-O3b-NSBH, popsyn-nsbh-Pawel-EjectaStudy2020, 2020ApJ...899L...1Z,2020ApJ...901L..39O,
    broekgaarden2021formation, broekgaarden2021impact}.
The population of gravitational-wave signals from coalescing compact binaries
    has an immediate impact on our understanding of the evolutionary tracks
    available to massive stellar binaries,
    even with the modest assumption that binary stellar astrophysical processes
    dominate at least some key parts of the observed compact binary distribution.

The formation channels for compact binaries include the 
    isolated evolution of massive stellar binaries,
    isolated stellar triples,
    and dynamic mergers in dense environments such as
    near galactic nuclei, globular clusters, and dense star clusters
    (see Mandel \& Farmer \cite{MANDEL20221} for a review).
Several theories postulate
    that many of the mergers observed thus far by modern ground-based
    gravitational-wave detectors can be explained by isolated binary
    evolution.
    \citeIsolated.
Others have argued that dynamic mergers are necessary to explain
    some events (like GW190521)
    \cite{Gayathri2022,Gamba2021,LIGO-O3-O3a-RP,LIGO-O3-O3b-RP, RomeroShawGW190521, Fragione_2020}.
In this work we assume that the collective population of compact
    object mergers was formed through isolated binary evolution
    and reserve the inclusion of dynamic channels for future work.

Many groups have undertaken the pertinent task of modeling and simulating
    the evolution of massive isolated stellar binaries
    to construct simulated populations of mergers to compare with
    observations \citeBE.
Previous studies with synthetic data have suggested direct comparisons 
    with binary evolution catalogs
    could constrain binary evolution model parameters \cite{Barrett2017}.
These models depend on assumptions about the life and evolution of
    stellar binaries which vary from group to group and simulation to simulation,
    and include (but are not limited to) supernova shock propagation,
    pair instability, stellar wind, mass transfer, and metallicity.
In this work, we will refer to the specific binary evolution model parameters
    varied from one simulation to the next as formation parameters,
    $\FormationParameters$.
Comparing any individual, detailed model for compact binary formation
    to observations is straightforward:
    the number of detections and properties of each event seen in a particular
    GW survey are compared quantitatively to the population of simulated mergers
    for a given model of binary evolution by a conventional
    (inhomogeneous Poisson process, marginalized) likelihood
    \cite{Stevenson2015popsyn,Wysocki2018,LIGO-O2-RP,
        Smith2020popsyn,Roulet2020popsyn}.
Simulating these populations incurs a high computational cost,
    severely limiting the ability to thoroughly explore the parameter space.
In limited cases, pioneering studies have used post-processing to implement
    single low-dimensional models that vary a handful of parameters at a time,
    for example changing the spin distribution
    \cite{Wysocki2018} or modifying the relative proportions
    (``mixing fractions'')
    between two fiducial reference models
    \cite{Zevin2017popsyn,Bouffanais2019popsyn}.
To surmount this challenge, some groups have created surrogate models for costly
    models like binary star evolution
    \cite{Stevenson2015popsyn,Taylor2018popsyn,Wong2019popsyn},
    allowing them to make continuous predictions for some
    parameter distributions as a function of some
    formation parameters.

Recently, a some groups have begun to undertake the task
    of considering more than one set of model assumptions at once
    \cite{Stevenson2022}, however they stop short of doing inference
    to interpolate between costly simulations and select
    new model parameter choices for further simulations.
In practice, however, these complex models have many parameters and are
    undergoing tremendous developments that 
    can dramatically impact their predictions,
    such as the choice for supernova engine, remnant spins, and common-envelope assumptions.
It would be difficult for these approximations to stay current and
    incorporate all relevant model physics or parameters.

In this work,
    we introduce a strategy to allow inference on compact binary formation models,
    without requiring continuously sampled models.
Motivated by highly-successfully strategies to interpret individual GW observations
    \cite{RIFT},
    we propose interpolating the marginalized likelihood over formation parameters
    and performing a Bayesian inference in these formation parameters.
In this study, we vary formation parameters for \kicks ($\STkick$),
    mass transfer efficiency during Roche-lobe overflow ($\STFa$),
    specific angular momentum of material ejected during
    Roche-lobe overflow ($\STBeta$) \cite{Podsiadlowski1992}, and a reduction in wind mass-loss rates
    for hydrogen-dominated stars ($\STWindOne$).
    Here, $\STkick$ is the dispersion parameter of a Maxwellian distribution.
We compare GW observations from the first
    three observing runs of the LIGO/Virgo instruments
    \cite{GWTC-1,GWTC-2,GWTC-2p1,GWTC-3}
    to models produced by the \startrack{} 
    binary evolution suite.
Coupled with a range of post-processing tools, the \startrack{} suite produces merger rate densities for
    a population of compact binaries formed over cosmological time,
    accounting for a distribution of star forming conditions
    at each formation redshift
    \cite{DominikI,DominikII,Belczynski2010metallicity,Belczynski2016Nature,Belczynski2020-EvolutionaryRoads,Belczynski2008}.

Astrophysical interpretation relies on the joint likelihood
    $\pprob (\gwdets|\FormationParameters)$
    of some specific model (with parameters $\FormationParameters$)
    given the data ($\gwdets$, consisting of individual confident detections: $\gwdetj$).
The Poisson likelihood, selection biases, observation results, and astrophysical inputs needed to carry out
    this program are readily available and well-understood
    \cite{LIGO-T1500562,LIGO-P1600187,LIGO-T1600208,Vitale_2021}.
Several groups have already demonstrated how to use the likelihood of
    individual discrete models to discriminate between them
    \cite{Wysocki2018,Zevin2017popsyn,Stevenson2015popsyn}.
However, at present this approach remains tightly limited by model availability:
    the event rate as a function of formation parameters
    is only available for a very small set of population model parameters.

We organize the paper as follows.

Section \ref{sec:methods} reviews all the methods used in this work.
First, in Section \ref{sec:startrack}, we review the \startrack{}
    binary evolution code,
    our binary evolution assumptions,
    and the range of binary evolution parameters
    explored by models used for inference in this study.

Next, in Section \ref{sec:populations},
    we review how we incorporate GW detector sensitivity to assess the properties
    of observed populations,
    carefully describing how our calculations use the \startrack{} binary evolution code.
We specifically describe the fiducial detector network sensitivity estimate
    (i.e. the adopted detector noise power spectrum and signal-to-noise threshold for detection)
    adopted for the rest of this work.

Subsequently, in Section \ref{sec:inference},
    we describe how we evaluate the marginal likelihood of each model,
    as compared to gravitational-wave observations.

Section \ref{sec:kick} demonstrates our approach to binary population inference,
    in the context of a simple one-parameter investigation of
    \kick{} velocity.
This section and the next compare observations from the first three observing runs
    of the LIGO/Virgo instruments to population models.

Section \ref{sec:results-full} applies these techniques to draw conclusions on
    the four formation parameters we explore in this study:
    $\STFa$, $\STBeta$, $\STkick$, and $\STWindOne$.
Finally, section \ref{sec:conclusions} discusses the implications of our
    results in the field of binary evolution population synthesis.

\section{Methods}
\label{sec:methods}
\begin{subsection}{\startrack{} Simulated Universes}
\label{sec:startrack}

\startrack{} binary evolution results have been exhaustively discussed before,
    both in isolation
    \cite{DominikI,DominikII,DominikIII,
        Belczynski2016,Belczynski2010metallicity,
        Belczynski2016Nature,Belczynski2016PSN,
        Wiktorowicz2019},
    and in comparison with GW observations
    \cite{Wysocki2018,Belczynski2016Nature,Belczynski2016,Belczynski2020b,
        Belczynski2020-EvolutionaryRoads,
        popsyn-nsbh-Pawel-EjectaStudy2020}.
O'Shaughnessy et al. \cite{ROSPSconstraints, ROSPSmoreconstraints}
    have systematically randomly varied many binary
    evolution parameters simultaneously,
    comparing them against an observed sample (of binary pulsars).
While this previous study also used a Bayesian,
    single-event-likelihood-based approach to assess the likelihood
    of given population models,
    our investigation is the first to carefully interpolate between 
    choices of formation parameters
    (see, however, the thesis work of Delfavero \cite{DelfaveroDissertation}).
The content in this section is based on
this prior work
    whose pertinent results we summarize for self-completeness.

\begin{subsubsection}{\startrack{} Simulations}
Synthetic universes are generated from a sequence of distinct \startrack{} runs,
    applied to a fixed number of isolated binaries all born at a fixed reference time.
For each set of binary evolution parameters,
    \startrack{} is applied to a range of progenitor metallicities.
With each run, \startrack{} evolves pairs of stars,
    accounting for accretion, tidal interactions,
    stellar wind, metallicity,
    gravitational radiation,
    magnetic braking,
    compact object recoil kicks,
    pair production instability,
    and many more physical processes.
It uses phenomenological models for supernovae to determine the properties of
    remnant black holes and neutron stars
    \cite{Fryer2012, Belczynski2010metallicity}.
Among its many outputs are the expected population of compact binaries --
    \bbh{}, \bns{}, and \nsbh{} systems --
    characterized as (weighted) samples from the distribution of merging binaries.

The initial population of stars in a \startrack{} simulation is drawn from
    a Kroupa initial mass function \cite{Salpeter, Kroupa1993, Kroupa2003}.
\begin{align}\label{eq:IMF}
\Psi(m_1) \propto 
\begin{Bmatrix}
    m^{-1.3}_1 & 0.08 \msun \leq m_1 < 0.5 \msun \\
    m^{-2.2}_1 & 0.5 \msun \leq m_1 < 1.0 \msun \\
    m^{-\alpha_{IMF}}_1 & 1.0 \msun < 150 \msun
\end{Bmatrix}
\end{align}
where we adopt $\alpha_{IMF} = 2.35$ for the highest mass stars
    (consistent with \cite{Belczynski2020-EvolutionaryRoads,Klencki2018}).
The companion mass ($m_2$) is drawn uniformly in mass ratio ($m_2/m_1$) from
    $[q_{\mathrm{min}}, 1]$
    where $q_{\mathrm{min}} = 0.08 \msun/m_1$ represents the hydrogen burning limit for $m_2$
    \cite{kobulnicky, Bastian2010, Duchene2013Review}.
For computational efficiency,
    only binaries where $m_1 > 5 \msun$ and $m_2 > 3 \msun$ are evolved
    when simulating \startrack{} populations of compact binaries,
    as only these systems can form black holes and neutron stars.

The first step in relating these \startrack{} runs to a synthetic universe is
    describing them as the population of stars produced by a specific
    amount of star-forming mass.
In other words, we compute the amount of star-forming gas $M_{\mathrm{sim}}$
    that would be expected to produce the population of evolved binaries,
    accounting for the arbitrary thresholds used to improve computational
    efficiency.
This is done by finding the mass efficiency, $\lambda_{\mathrm{sim}}$ \cite{2010binary},
    where
\begin{align}\label{eq:mass-frac}
\lambda = \frac{n}{N} \frac{f_{\mathrm{cut}}}{\langle M \rangle}
\end{align}
where $N$ is the number of total binaries simulated and $n$ is the number
    of compact binary progenitors,
    $\langle M \rangle$ is the average mass of all binary progenitors,
    and $f_{\mathrm{cut}}$ accounts for the binaries that are cut
    to fit mass requirements.

$M_{\mathrm{sim}}$ must also account for systems which do not form binaries,
    systems which do not merge in the Hubble time,
    and a sea of low mass stars.
While the true fraction of stars which form binaries is unknown,
    the merger rate ($\rho(t|f_b)$) per unit mass at time $t$ for any choice of binary fraction
    $f_b$, can be found in terms of the corresponding answer for $f_b=1$
    by using the following relationship \cite{2010binary}:
\begin{align} \label{binary-frac}
\rho(t|f_b) = \rho (t|f_b = 1)
    \frac{f_b(1 + \langle q \rangle)}{1 + f_b \langle q \rangle}
\end{align}
    where $\langle q \rangle = 0.5$ is the expected mass ratio
    (for uniform $q$; consistent with Sana et al 2012 \cite{Sana2012}),
    and $\rho (t | f_b = 1)$ is the merger rate versus
    cosmological time evaluated for $f_b=1$.
$f_b$ is assumed to be $1/2$ unless otherwise noted.

One can find $M_{\mathrm{sim}}$, the mass representing a \startrack{}
    simulated population, as:
\begin{align}\label{eq:msim}
M_{\mathrm{sim}} = \frac{2f_b}{1 + f_b} \frac{n}{N} \frac{f_{\mathrm{cut}}}{\langle M \rangle} M_{\mathrm{bin}}
\end{align}
where $M_{\mathrm{bin}}$ is the sum of the mass of all simulated binaries.

The output of \startrack{} is a list of simulated compact binary mergers
    and their binary parameters, $\BinaryParameters$,
    for a single Zero Age Main Sequence (ZAMS) metallicity, $\metallicity$,
    and a single set of binary evolution model constraints (
    i.e.
    formation parameters),
    $\FormationParameters$.
The binary parameters $\BinaryParameters$ include $m_1$, $m_2$, and redshift at time of merger ($\redshift_m$).
Each simulation provides an estimate for how densely the simulation
    is populated in the space of $\BinaryParameters$.
\begin{align}\label{eq:merger_rate_density}
\rho_{\mathrm{sim}}(\BinaryParameters) = \sum\limits_i \delta (\BinaryParameters - \BinaryParameters_i)
\end{align}
where $i$ indexes the coalescing binaries in that simulation.
This expression is normalized as:
\begin{align}\label{eq:merger_rate_bin}
N_{\mathrm{sim}} = \int\limits_{\AllBinaryParameters} \rho_{\mathrm{sim}}(\BinaryParameters)
    \mathrm{d} \BinaryParameters
\end{align}
Though not used in this work, 
    Eq. \ref{eq:merger_rate_density} and Eq. \ref{eq:merger_rate_bin}
    are conceptually helpful when connecting simulation
    Monte Carlo results to the explicit cosmological post-processing performed below.

\end{subsubsection}

\begin{subsubsection}{Single Star and Binary Evolution Assumptions}
\label{sec:models}

Beyond assuming an initial mass function,
    and bookkeeping with regard to how sample mergers
    represent a simulation mass,
    there are many assumptions about individual and binary stars
    that deserve accounting.
Exhaustively describing every assumption used by \startrack{}
    is unnecessary, as \startrack{}
    has been developed alongside literature
    \cite{Belczynski2008, Belczynski2016Nature, Belczynski2020-EvolutionaryRoads}.
However, in order to effectively compare assumptions made in this work
    to those of other groups currently studying binary evolution,
    I will review some of the most relevant assumptions.

\paragraph{Supernova Engine}

The models M13, M14, M15, M16, M17, M18, and M19
    are based on the recently studied M10 model
    \cite{Belczynski2020-EvolutionaryRoads},
    and differ in assumptions about kick velocity.
For the models based on the M10 model
    SN masses are drawn from the ``rapid'' SN engine of Fryer et al 2012
    \cite{Fryer2012},
    considering
    neutrino mass loss at compact object formation.
For models based on M10,
    mass loss from neutrinos is 10 percent,
    regardless of source properties.
These models 
    strong effects from pulsational pair-instability supernova (PPISN)
    and strong pair-instability supernova (PISN),
    consistent with Belczynski et al. \cite{Belczynski2016PSN}.
These models assume a 10 percent Bondi-Hoyle accretion rate
    onto \nsbh{} during CE events.
Remnant black hole spins for models based on M10
    are computed using Geneva stellar evolution models,
    following \cite{Belczynski2020-EvolutionaryRoads}.

The K series of models (K0100-K0598)
    are derived from the M30 model
    \cite{Belczynski2020-EvolutionaryRoads}
    rather than M10.
While the M30 model incorporated a ``rapid`` SN engine
    with weak PISN/PPISN effects 
    \cite{Fryer2012, Belczynski2016PSN},
    K0100-K0598 incorporate a ``delayed'' SN engine
    (see the same papers).
Mass loss from neutrinos in these models is divided
    based on source properties.
In massive black holes ($ > 3 M_{\odot}$),
    10 percent of the SN remnant mass is lost to neutrino emission,
    while 1 percent loss is assumed for less massive remnants.
These models assume a 5\% Bondi-Hoyle accretion rate
    onto \nsbh{} during CE events.
Remnant black hole spins for models based on M30
    are based on MESA simulations
    \cite{Belczynski2020-EvolutionaryRoads}.

All models assume a maximum neutron star mass of 2.5 $M_{\odot}$.
Rather than assuming a fixed \bbh{} efficiency,
    systems are evolved and the end of their evolution is determined
    by the properties of each system.

\paragraph{Supernova Recoil Kicks and Fallback}

We note that while the M10 model does have a reduction in kick velocity
    due to matter falling back onto a new compact remnant,
    but M13-M19 do not.
Unless otherwise noted,
    \kick velocities for new compact remnants are drawn from a 
    single-peaked Maxwellian distribution
    characterized by the dispersion 
    $\STkick$ (with units km/s).
This velocity is not reduced by fallback.

A one-parameter family of models varying this dispersion
    (between 10. and 265 km/s)
    has been previously presented
    \cite{Belczynski2016Nature, Belczynski2016PSN}
    and compared to GW observations in
    \cite{Wysocki2018}.
These models are denoted M13, M14, M15, M16, M17, and M18 (as well as M19 which
    was generated for this publication).
This parameter is also varied in our higher-dimensional studies.

The K series models based on M30 presented in this publication
    are described by the same kick model as M13-19.

Other, more complex \kick{} models exist
    and have been studied in other work
    \citeBetterKicks{}.
It is important to keep in mind 
    the uncertainty introduced by this simplified kick model
    when interpreting our results.
For a discussion of the mechanisms involved with 
    both fallback-reduced kicks and 
    non-fallback-reduced kicks,
    see Section 6.2 of \cite{Belczynski2016}.

\paragraph{Mass Transfer}

Roche-lobe overflow is the transfer of mass from one star (the donor)
    to its companion (the accretor) when the atmosphere of the donor expands 
    beyond the region where mass is gravitationally bound to just that star.
This transfer occurs near the point between two stars where their gravitational
    pull cancels out.
Roche-lobe overflow can be stable or unstable
    (e.g. if the donor's roche-lobe shrinks faster than its
    volume during mass transfer).
Instabilities can lead to a merger or common envelope phase.
Stable roche-lobe overflow can be conservative or non-conservative.
For (dynamically) stable non-conservative roche-lobe overflow,
    we adopt the formalism seen in
    Rappaport, Joss, and Webbink \cite{Rappaport1982},
    Rappaport, Verbunt, and Joss \cite{Rappaport1983}.
    and later \cite{Podsiadlowski1992}.

The fraction ($\STFa$) of mass lost by the donor which is accreted
    onto its companion,
    and therefore $1 - \STFa$ is the amount of matter
    which becomes unbound and is lost by the system.
Some angular momentum is lost by this ejected material,
    described by the equation:
\begin{align}
\delta J = \STBeta \delta \dot{M_1} (1 - \STFa) \frac{2 \pi a^2}{P}
\end{align}
where $\delta \dot{M_1}$ is an infinitesimal amount of mass lost
    by the donor, $\delta J$ is a small amount of angular momentum
    carried by matter ejected from the system,
    $P$ is the orbital period,
    and $a$ is the semi-major axis of the orbit.
This angular momentum loss to ejected material
    is constrained by an efficiency ($\STBeta$),
    which is also referred to as the specific angular momentum of the ejected material.

In our one-dimensional study,
    these parameters are fixed
    ($\STFa = 0.5$ and $\STBeta = 1.0$).
In our higher-dimensional study,
    we vary these parameters in order to demonstrate the effectiveness
    of observation-driven constraints on these parameters
    from gravitational-wave events.

For unstable roche-lobe overflow resulting in a common envelope,
    we adopt methodology consistent with \cite{Belczynski2020-EvolutionaryRoads}
    for common envelope evolution.
In the models presented by this work,
    common envelope efficiency (as often described by $\alpha_{\mathrm{CE}}$)
    is set equal to unity.
Some work has been done to study $\alpha_{\mathrm{CE}}$
    (e.g. \cite{Wilson2022, Klencki2021}),
    however there is still much that is not understood about common envelope evolution.
The common envelope binding energy (as often described by $\lambda_{\mathrm{CE}}$)
    is consistent with \cite{Belczynski2020-EvolutionaryRoads},
    which is adapted from \cite{XuLi2010} and \cite{IvanovaRemnant2011, IvanovaEnthalpy2011}.

\paragraph{Stellar Wind}
\label{sec:wind}

Consistent with other literature on \startrack{}
    \cite{Belczynski2010metallicity, Belczynski2020-EvolutionaryRoads},
    we start from standard Vink wind models \cite{Vink2001}.
However, inhomogeneities and clumping for line-driven winds
    from galactic \citeWeakWind{} and Small Magellanic Cloud (SMC)
    \citeWeakWindSMC{} sources
    have shown that Vink winds may overestimate mass-loss rates
    in massive stars.
Theoretical and observational studies
    of this ``weak wind phenomenon''
    \citeWeakWindFactor{},
    suggest that mass-loss rates may be overestimated
    by a factor of about three.

For models discussed in Section \ref{sec:results-full},
    we implement a scaling reduction
    in stellar wind mass-loss rates compared to the standard
    Vink model \cite{Vink2001}.
For a factor of three reduction,
    this scale factor ($\STWindOne$ for hydrogen-dominated stars)
    would have a value of $1/3$.
For helium-dominated stars, we leave this scale-factor ``off''
    ($\STWindTwo=1.0$).
This reduction in mass-loss rates due to stellar wind was introduced to
    \startrack{} by Belczynski et al \cite{Belczynski2020b}.
One of the major goals of this work is to 
    test how these lower winds impact properties of 
    the merger population.

\end{subsubsection}

\begin{subsubsection}{Cosmological Post-processing}\label{sec:postprocessing}    
\begin{figure*}
\centering
\includegraphics[width=\textwidth]{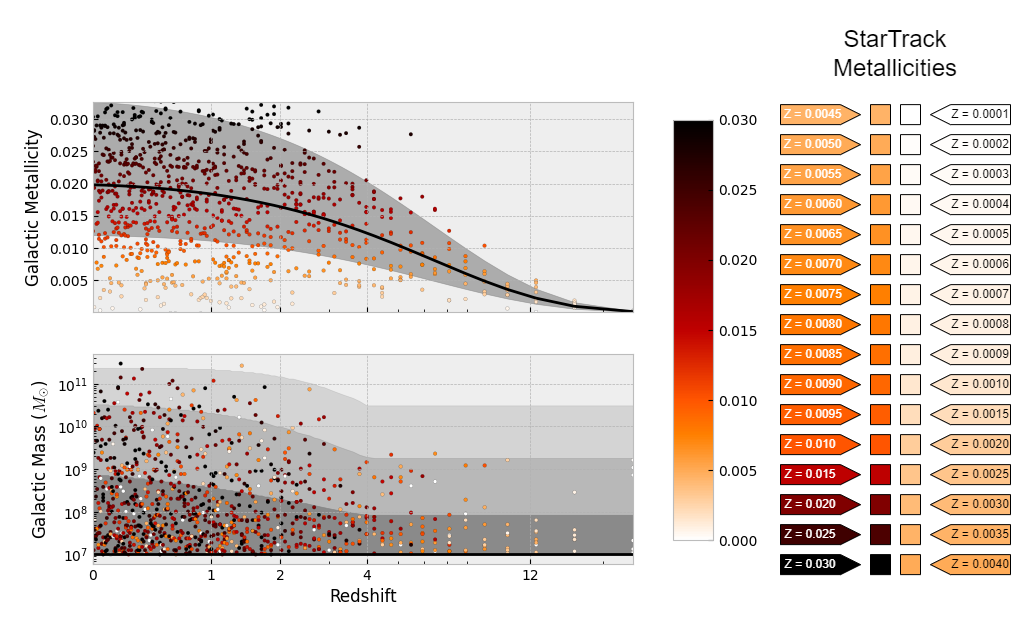}
\caption{\label{fig:Metallicity}
    \textbf{Metallicity and mass distributions of star forming material as a function of cosmological redshift:}
    During postprocessing, star formation in each epoch of the universe's history
        is represented by sampling galactic metallicity and star-forming mass distributions,
        assigning each galaxy to
    the
    nearest \startrack{} metallicity bin
        and scaling \startrack{} simulation masses by the mass of each galaxy.
    This enables a physically motivated accounting of star formation
        in each epoch of the universe's history.
    (top-left): 
        The Madau and Fragos \cite{Madau2017}
        metallicity dependence on redshift assumed in this work,
        where the black line follows 
        Eq. (6) of \cite{Madau2017}.
        Samples (colored by metallicity)
            are drawn from the Cumulative Distribution Function (CDF)
            of the one-dimensional Gaussian with an assumed uncertainty of
            0.5 dex (gray shaded region),
            consistent with prior work
            \cite{Belczynski2020-EvolutionaryRoads}.
    (bottom-left):
        The Madau and Fragos \cite{Madau2017} galactic star forming mass
            dependence on redshift (
            see Eq. (1) of \cite{Madau2017}
        ).
        Shaded regions represent the $68$, $95$, and $99.7\%$ confidence regions.
        Samples are again colored by metallicity.
    (right):
        The \startrack{} metallicity bins used in this study.
        Each sample is assigned to a bin most closely associated with
            its given metallicity.
}
\end{figure*}

The \startrack{} team has developed routines to generate
    synthetic universes from an underlying set of \startrack{} simulations
    \cite{DominikII,DominikIII,Belczynski2020-EvolutionaryRoads}.
These post-processing routines draw samples from the \startrack{} simulations,
    assigning each simulated binary a weight which represents
    the expected merger rate in each simulated universe.
This method of drawing from multiple simulations at different ZAMS metallicities
    yields a distribution which can be compared to a particular epoch of star
    formation in the history of the universe.

In this work, we follow the procedures described in Section 2.6 of
    \cite{Belczynski2020-EvolutionaryRoads}
    in order to describe our choices of Star Formation Rate Density (SFRD),
    Galaxy Stellar Mass Function (GSMF)
    and Mass-metallicity Relation (MZR).
We note that these choices can have a dramatic impact on the
    predicted rates of compact binary mergers
    (as explored by \cite{broekgaarden2021impact}).
Future work may look further to explore these choices.
Specifically, we adopt the Madau \& Fragos (2017)
    \cite{Madau2017} SFRD,
    with an IMF-dependent correction factor;
    see Eq. (1) of \cite{Madau2017}.

Also following from Madau \& Fragos (2017)
    \cite{Madau2017}
    (and subsequently \cite{DominikII} and \cite{Belczynski2020-EvolutionaryRoads},
    we adopt a mass-metallicity relation (MZR) so the average metallicity versus redshift
    satisfies
    Eq. (6) of \cite{Madau2017} (which is inspired by \cite{Zahid2014}):
\begin{align}
\label{eq:z_of_z}
    \log Z/Z_{\odot} = 0.153 - 0.074 \redshift^{1.34}
\end{align}
This mass-metallicity relation is enforced by dividing the history
    of the universe into $\Delta t = 100 \mathrm{Myr}$ epochs
    and inferring a mass distribution informed by redshift.
Each epoch of evolutionary history is characterized by
    a distribution of galaxy masses (GSMF).
We adopt a Schechter-type \cite{schechter} GSMF from \cite{fontana2006},
    frozen beyond redshift $\redshift = 4$.
    (consistent with \cite{DominikII}).

Rather than assigning every galaxy in a given epoch the same
    metallicity,
    galaxy samples are assigned a metallicity
    by assuming a Gaussian
    where the mean metallicity ($\bar{\metallicity}$)
    at a given redshift is assumed from Eq. \ref{eq:z_of_z}
    and $\sigma \metallicity = \exp(-0.5) \bar{\metallicity}$
    (i.e. 0.5 dex).
The percent point function (inverse of cumulative distribution function)
    of the Gaussian is sampled uniformly in a space truncated
    on either side by eleven standard deviations.

We adopt a solar metallicity as $Z_{\odot} = 0.02$.
As there is not a \startrack{} simulation for every possible metallicity value,
    these galaxies are gathered into metallicity ``bins'' representing the closest
    metallicity value for which there is a \startrack{} simulation.
The metallicity bins included in our study are:
$\metallicity \in \{$
    $0.0001$, $0.0002$, $0.0003$, $0.0004$, $0.0005$, $0.0006$, $0.0007$, $0.0008$, $0.0009$,
    $0.001$, $0.0015$, $0.002$, $0.0025$, $0.003$, $0.0035$, $0.004$, $0.0045$, $0.005$,
    $0.0055$, $0.006$,
    $0.0065$, $0.007$, $0.0075$, $0.008$, $0.0085$, $0009$, $0.0095$, $0.01$, $0.015$,
    $0.02$, $0.025$, $0.03\}$ \cite{Belczynski2020-EvolutionaryRoads}.
Figure \ref{fig:Metallicity} depicts this sampling,
    from the Madau \& Fragos models \cite{Madau2017}.

Although each metallicity bin requires a separate \startrack{} simulation,
    each time bin does not,
    as it is only used to construct a distribution in 
    the ZAMS metallicity of sources at a given time in the history of the universe.
The weight function that we need to predict compact binary formation
    is now simply
\begin{align}\label{eq:f_r}
\rho_{\Lambda, \metallicity, \Delta t} (\BinaryParameters)
    = \frac{M_{\FormationParameters, \metallicity, \Delta t}}{M_{\FormationParameters,
    \{\metallicity\}, \Delta t}}
    \frac{\mathrm{SFR}(\redshift)}{M_{\mathrm{sim}}}
\end{align}
where $\mathrm{SFR}(t)$ is the cosmological star formation rate,
    for time bins which ultimately correspond to a redshift.
Here, $M_{\FormationParameters, \metallicity, \Delta t}$ is the mass attributed
    to a \startrack{} simulation with formation parameters
    $\FormationParameters$, and metallicity $\metallicity$,
    for a particular epoch of time $\Delta t $.
Additionally, $M_{\FormationParameters, \{\metallicity\}, \Delta t}$ is the mass
    attributed to a particular epoch of time, $\Delta t$,
    for a particular \startrack{} simulation with parameters $\FormationParameters$

These weights are assigned to mergers in the output of
    \startrack{} for each metallicity bin,
    and the sample population is now weighted correctly
    to represent the physical universe.
Following this, the merger density for a synthetic universe is:
\begin{align}\label{eq:s_i}
\rho_{\FormationParameters}(\BinaryParameters) = \
    \sum\limits_{\Delta t, \metallicity}
        \rho_{\FormationParameters, \metallicity, \Delta t}(\BinaryParameters)
        \Delta t \Delta \metallicity
\end{align}

Throughout the rest of this paper,
    $\rho_{\FormationParameters}(\BinaryParameters)$ refers
    to the merger rate density for a synthetic universe
    constrained by the formation parameters $\FormationParameters$
    (with units of $\mathrm{Mpc}^{-3} \mathrm{yr^{-1}} [\BinaryParameters]^-1$).
This density is ultimately composed of the discrete merger samples:
\begin{align}\label{eq:weighted_density}
\rho(\BinaryParameters) = \sum\limits_i s_i \delta (\BinaryParameters - \BinaryParameters_i)
\end{align}
where $s_i$ is the weight given to a particular sample merger.
\end{subsubsection}
\end{subsection}

 \begin{subsection}{Relating Detected and Observed Populations}
\label{sec:populations}

Our single-detector model for network sensitivity relates the populations
    of compact binary mergers assembled by our simulations
    to the observed population of gravitational-wave events.
In order to perform that calculation
    we make use of the merger samples and their weights
    to perform an integral over the sample population.
The work in this section follows \cite{DominikIII}.

The expected gravitational-wave detection rate for a sample
    of the merger population from a given synthetic universe,
    characterized by formation parameters $\FormationParameters$,
    (compare to Eq. 5 of \cite{DominikIII}) is given by:
\begin{align}\label{eq:R_det}
R_{\FormationParameters}(\BinaryParameters) = \rho(\BinaryParameters)
    p_{\mathrm{det}}(\BinaryParameters)
    \frac{\mathrm{d}V_c}{\mathrm{d}\redshift_m}
    \frac{\mathrm{d}t_m}{\mathrm{d}t_{\mathrm{det}}}
\end{align}
Here, $\frac{\mathrm{d}t_m}{\mathrm{d}t_{\mathrm{det}}} =
    \frac{1}{1 + \redshift}$
    is the factor relating merger time and detector time.
$p_{\mathrm{det}}$ is the detection probability (see Eq. \ref{eq:pdet}).
Eq 6 of \cite{DominikIII} describes the differential comoving volume:
\begin{align}\label{eq:tides_6}
\frac{\mathrm{d}V_c}{\mathrm{d\redshift}} =
    \frac{4 \pi c}{H_0} \frac{D^2_c (\redshift)}{E(\redshift)}
\end{align}
with comoving distance, $D(\redshift)$, and the dimensionless
    cosmological scale factor, $E(\redshift)$ \cite{DominikIII}.
In our calculations we use the conventional Planck2015 cosmology
    \cite{Planck2015},
    implemented through astropy for cosmological quantities
    such as $\mathrm{d}V_c/\mathrm{d}\redshift$
    \cite{astropy:2013,astropy:2018}.

The total number of gravitational-wave detections of a given kind
    ($\alpha \in \{\mathrm{\bbh{}, \bns{}, \nsbh{}}\}$)
    expected of a simulated universe is given by:
\begin{align}\label{eq:detection-rate}
\mu_{\FormationParameters,\alpha} = T_{\mathrm{obs}}
    \iiint\limits_{0}^{\infty}
        R_{\FormationParameters,\alpha}(m_1, m_2, \redshift_m)
        \mathrm{d}m_1 \mathrm{d}m_2 \mathrm{d}\redshift_m
\end{align}
where other merger parameters $\BinaryParameters$ have been marginalized out.
$T_{\mathrm{obs}}$ is the observing time for a run
    of the gravitational-wave observatories (LIGO and/or Virgo).
Finally, types of mergers, $\alpha$, are distinguished by a simple mass threshold
    ($m > 2.5 \rightarrow \mathrm{BH}$; $m < 2.5 \rightarrow \mathrm{NS}$).

Performing this integral over all the sample mergers for a 
    \startrack{} synthetic universe with one set of model assumptions
    (the formation parameters, $\FormationParameters$)
    yields an expected number of 
    gravitational-wave observations from ground based observatories
    with those assumptions.
The density of these predicted detection rates for each synthetic universe
    can also be interpreted in one- and two-dimensional marginalizations,
    providing an insightful glimpse into the predicted population
    of compact binaries
    in a synthetic universe (see Figure \ref{fig:M15}).
By comparing these predicted detection rates and populations to observations,
    we will limit the range of formation parameters
    motivated by  Bayesian inference, described in greater detail
    in Section \ref{sec:inference}.

\begin{figure}
\centering
\includegraphics[width=3.375 in]{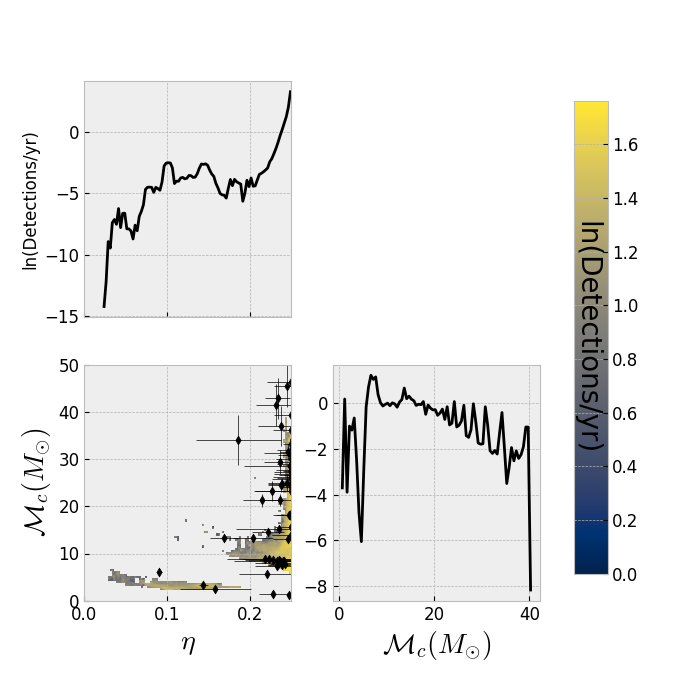}
\caption{\label{fig:M15}
\emph{Off-diagonal:}
    A two-dimensional histogram illustrates an example of the
    predicted mass distribution of detections
    ($\mathcal{R}_det (\mathcal{M}_c, \eta)$ --
    re-parameterized from $m_1$ and $m_2$;
    see Eq. \ref{eq:R_det})
    for the M15 model 
    (see Sections \ref{sec:models} and \ref{sec:kick} for more about M15).
Also plotted are MLE parameter values
    (see Figure 5 of \cite{nal-methods-paper})
    for the confident GW events
    discussed by the LIGO/Virgo/KAGRA collaborations in
    their rates and populations study following the third observing run
    of the ground-based gravitational-wave observatories
    \cite{LIGO-O3-O3b-RP}.
\emph{Diagonal:} one-dimensional detection-weighted densities
    for predicted detections in M15 in $\mathcal{M}_c$ and $\eta$.
Notice the distinct lack of predicted events in the lower mass gap,
    and of high-mass events in the 40-50 $\msun$ range.
}
\end{figure}

\begin{subsubsection}{Detection Probability}
For interferometers like LIGO and Virgo,
    the detection probability depends strongly on the properties of each source.
The process of detection is a complex process,
    depending on search pipelines and data
    quality vetoes which involve many human choices 
    \cite{GW150914-detection,CalibrationPaper,GWTC-1,O3-Detector,Virgo}.
We cannot hope to reproduce this full process across the many potential
    observing runs each of our simulations can produce.
We make a standard assumption that our survey can be approximated
    by a single interferometer,
    with a detection being defined as producing a signal-to-noise ratio (SNR)
    greater than a threshold,
    which we choose as $\rho_{\mathrm{thr}} = 8$.
Using this approximation and averaging over all source orientations, 
    the detection probability for a source with
    specific intrinsic parameters and redshift ($\BinaryParameters$) 
    but averaged over the sky and all source orientations
    can be expressed as a detection \emph{probability} $p_{det}(\BinaryParameters)$.
When evaluating this expression,
    we need the SNR for an optimally
    located and oriented binary with parameters $\BinaryParameters$,
    which in turn depends on our assumptions about our fiducial
    detector's sensitivity. 

From the O2 catalog, we see that the \bns{} range for
    the LIGO Hanford Observatory (LHO) and LIGO Livingston Observatory (LLO) was up to
    80 Mpc.\cite{GWTC-1}
As in previous work,
    this corresponds to a high sensitivity Point Spread Distribution (PSD)
    for early LIGO observation runs 
    (the ``SimNoisePSDaLIGOEarlyHighSensitivityP1200087''
        PSD included in lalsuite)\cite{P1200087,lalsuite}.
For the first/second part of O3,
    we see that the \bns{} range for LHO is 135/133 Mpc,
  and is 108/115 Mpc for LLO
  \cite{GWTC-2,GWTC-3}.
The sensitivity of Virgo is not considered in our
    single-detector model.
For O3, we use an optimistic model with the
    ``SimNoisePSDaLIGOaLIGO140MpcT1800545'' PSD
    included in lalsuite;
    this allows us to include a correction factor to account for the
    VT estimate for O3 \cite{LIGO-O3-O3b-RP, lalsuite, P1200087, T1800427}.

In our work, 
    the SNR is calculated for each simulated binary in every synthetic universe
    (trained using the lalsuite optimal matched-filter SNR
    for each PSD mentioned above;
    see Appendix \ref{ap:SNR})
    \cite{lalsuite, popmodels}.
We also define a threshold SNR, $\snr_{\mathrm{thr}}=8$,
    above which the detection is counted,.
Following Dominik et al. \cite{DominikIII}, we find the probability of detection:
\begin{align}\label{eq:pdet}
    p_{det}(\lambda) = P\Big(\frac{\snr_{thr}}{\snr(\lambda)}\Big)
\end{align}
where $\pprob(w)$ is interpolated from results
    tabulated by other groups (https://pages.jh.edu/\~eberti2/research/)
    \cite{DominikIII}.
\end{subsubsection}

\end{subsection}
 \begin{subsection}{Bayesian Inference for Populations}
\label{sec:inference}

In the previous sections,
    we discussed our methods for using \startrack{} simulations
    to predict compact binary merger rates for a simulated universe
    and calculating predicted gravitational-wave detection rates associated
    with those merger densities.
Now, we set about using those tools to draw conclusions
    about The universe, expanding on the Bayesian framework
    described by \cite{DominikIII,Wysocki2018,Wysocki2019,DelfaveroDissertation}.

The predicted gravitational-wave detections, $\gwdets$,
    and merger population,
     $\rho_{\FormationParameters,\alpha}$,
    can be compared with a set of
    observations ($\gwdets$) using an inhomogeneous Poisson point process
    \cite{Stevenson2015popsyn,Wysocki2018,LIGO-O2-RP,Smith2020popsyn,Roulet2020popsyn}.
The likelihood associated with this process can be broken down:
\begin{align}\label{eq:joint-likelihood-simple}
    \pprob(\gwdets|\FormationParameters) =
        \pprob(\mu|\FormationParameters) \prod\limits_{j} \pprob(\gwdetj|\FormationParameters)
\end{align}
    where $\mu$ is the number of \emph{observed} GW detections
    and $\pprob(\gwdets|\FormationParameters)$ is ultimately the quantity used
    to evaluate the agreement between observations and 
    the predicted detection rates and merger populations for a simulation
    ($\mu_{\FormationParameters,\alpha}$ and $\rho_{\FormationParameters}$, 
    for kind $\alpha \in \{\mathrm{\bbh{}}, \mathrm{\bns{}}, \mathrm{\nsbh{}}\}$).
Throughout this work,
    we refer to $\pprob(\gwdets|\FormationParameters)$ as the ``joint likelihood'',
    $\pprob(\mu|\FormationParameters)$ as the ``rate likelihood'',
    and $\prod\limits_{j} \pprob(\gwdetj|\FormationParameters)$ as the
    ``shape likelihood.''

\begin{subsubsection}{Rate Likelihood} \label{sec:rate-likelihood}
The \emph{rate likelihood} is calculated the same as a standard Poisson point process
    for each type of event:
\begin{align}\label{eq:poisson-likelihood}
\pprob(\mu_{\alpha}|\FormationParameters) =
    e^{-\mu_{\FormationParameters,\alpha}}
    \frac{\mu_{\FormationParameters}^{N_{\alpha}}}{N_{\alpha}!}
\end{align}
where $N_{\alpha}$ is the number of observed GW sources of a given type.
The total rate likelihood is the product of these:
\begin{align}\label{eq:rate-likelihood}
\pprob(\mu | \FormationParameters) =
    \pprob(\mu_{\mathrm{\bbh{}}} |\FormationParameters)
    \pprob(\mu_{\mathrm{\bns{}}} |\FormationParameters)
    \pprob(\mu_{\mathrm{\nsbh{}}} |\FormationParameters)
\end{align}
For a breakdown of how to efficiently calculate this quantity directly in the log
    space, see Section 4.1.1 of \cite{DelfaveroDissertation}.

\end{subsubsection}
\begin{subsubsection}{Shape and Joint Likelihood}\label{sec:shape-likelihood}
The \emph{shape likelihood} of a synthetic universe with formation parameters, $\FormationParameters$,
    describes the probability that each individual GW source could be produced in a universe
    where those assumptions are true.
More specifically, it measures the agreement of the samples described by 
    $\rho_{\FormationParameters}(\BinaryParameters)$ to the shape of the likelihood function
    for the binary parameters of each detection, $\mathcal{L}_j(\BinaryParameters)$.
This likelihood is marginalized over the merger samples:
\begin{align}\label{eq:shape-likelihood}
    \pprob(\gwdetj|\FormationParameters) =
    \int\limits_{\AllBinaryParameters}
        \pprob(\gwdetj|\BinaryParameters,\FormationParameters)
        \pprob(\BinaryParameters|\FormationParameters)
        \mathrm{d}\BinaryParameters
    = \int\limits_{\AllBinaryParameters}
        \bar{\rho}_{\FormationParameters}(\BinaryParameters)
        \mathcal{L}_j(\BinaryParameters)
        \mathrm{d}\BinaryParameters
\end{align}
where $\bar{\rho}_{\FormationParameters}(\BinaryParameters) =
    \rho_{\FormationParameters}(\BinaryParameters) /
    \int_{\AllBinaryParameters} \rho_{\FormationParameters}(\BinaryParameters ')
    \mathrm{d} \BinaryParameters'$.

This marginalization is carried out over the entire population of
    $~10^8$ sample mergers in a synthetic universe,
    for each GW observation from a given observing run.
We make use of bounded (truncated) multivariate normal distributions to calculate this likelihood,
    described in separate publications 
    \cite{nal-chieff-paper,nal-methods-paper,DelfaveroDissertation}.
While the estimated likelihood for individual gravitational-wave
    events can have more complex morphology,
    these bounded multivariate normal distributions have been
    demonstrated to lose less information than waveform systematics
    \cite{nal-methods-paper}.
As relativistic waveforms are most sensitive to chirp mass and symmetric mass ratio
    and Gaussian noise is expected in these coordinates,
    these coordinates (in the source frame of reference)
    are used for evaluating this likelihood.
Putting this together with our rate likelihood,
    the ``joint likelihood'' can be expressed
    (equivalent to Eq. 4 of \cite{Wysocki2019}).
\begin{align}\label{joint-likelihood}
\pprob(\gwdets | \FormationParameters) =
    K_{\mathrm{rate}}
    e^{-\mu_{\FormationParameters}}
    \prod\limits_j \bigg[ \int\limits_{\AllBinaryParameters}
        \mathcal{L}_j(\BinaryParameters)
        \bar{\rho}_{\FormationParameters}(\BinaryParameters)
        \mathrm{d}\BinaryParameters \bigg]
\end{align}
where 
\begin{align}
K_{\mathrm{rate}} = 
    \frac{\mu_{\FormationParameters}^{N_{\mathrm{\bbh{}}}}}{N_{\mathrm{\bbh{}}}!}
    \frac{\mu_{\FormationParameters}^{N_{\mathrm{\bns{}}}}}{N_{\mathrm{\bns{}}}!}
    \frac{\mu_{\FormationParameters}^{N_{\mathrm{\nsbh{}}}}}{N_{\mathrm{\nsbh{}}}!}
\end{align}.

\end{subsubsection}
\begin{subsubsection}{Model Interpolation and Posterior Generation}
\label{sec:interpolation}

Interpolating each type of detection rate and
    likelihood in the formation parameter space
    is valuable in one- and higher-dimensional
    studies (see Figures \ref{fig:Kick:Comparison} and 
    \ref{fig:full-parameter-likelihood} respectively).
We use Gaussian process regression to implement such
    interpolations, and briefly summarize this method:
For brevity and to be consistent with conventional notation,
    we denote $\BinaryParameters$ by $x$ and the quantity being fit by $y$.
In this approach, we estimate the expected value of $y(x)$ from data $x_*$ and values
    $y_*$ via
\begin{eqnarray}
\E{y(x)} = \sum\limits_{\gamma, \gamma'} k(x,x_{*,\gamma})
    (K^{-1})_{\gamma,\gamma'} y_{*,\gamma'}
\end{eqnarray}
where $\gamma$ is an integer running over the number of training samples in $(x_*, y_*)$
    and where the matrix $K = k(x_{\gamma},x_{\gamma'}) y_*$.
The expected variance at $x$ is given by $K(x,x)^{-1}$.
We employ a kernel function $k(x,x')$ which allows for uncertainty in
    each estimated training point's value $y_{*,\gamma}$,
    reflecting systematic uncertainty in the input values.
We use a conventional piecewise-polynomial kernel to ensure compact support
    \cite{williams2006gaussian};
    the implementation of this
    kernel is discussed in the context of SNR in Appendix \ref{ap:SNR}.

One advantage of a Gaussian process is that it 
    can accept information about uncertainties in training data.
To account for systematic uncertainties in our binary evolution inputs,
    we adopt a fiducial (and optimistic) systematic uncertainty
    of $10$ percent
    in the event rate,
    and thus $0.1 \mu$ in the log-likelihood.
We emphasize our specific choice is an arbitrary division between nominally subdominant and dominant
  parameters (i.e., we assert parameters with less than 10\% effect are ignorable and account for them with systematic
  error).  Neither the specific nominal systematic error nor our limited model space are intended as definitive or even
  representative exploration of all possible parameters and uncertainties; rather, these choices allow us to illustrate
  the method of this paper with realistic models and assumptions, while allowing us to defer the exhaustive exploration
  of many more parameters to future work.When plotting our interpolated quantities,
    we show the nominal $1\sigma$ uncertainty predicted by the Gaussian Process.

We use two different methods for interpolating the joint likelihood:
(1) direct interpolation of $\pprob(\gwdets|\FormationParameters)$ for
    each simulation, and
(2) interpolation of $\mu_{\alpha}(\FormationParameters)$ for each type of
    merger, and interpolation of $\sum\limits_j \pprob(d_j | \FormationParameters)$
    (i.e. the shape likelihood).
    By interpolating these quantities, we can construct the joint likelihood
    with Eq. \ref{eq:joint-likelihood-simple} for each sample.

Of these, the first is more direct, and the second 
    is more useful in a sparsely sampled space.

Though sampling from a bounded multivariate normal distribution
    (i.e. truncated Gaussian) is used to propose new simulations
    for the models in Section \ref{sec:results-full},
    an interpolated model can be used for sampling when there is not
    a simple peak.
When sampling from an interpolated rate, shape, or joint likelihood
    to construct a posterior for that quantity,
    we sample uniformly in the formation parameter space,
    and find a weight for each point from the Gaussian process
    (assuming a uniform prior in $\FormationParameters$).

\end{subsubsection}
\begin{subsubsection}{Inputs from GW Observations}

The third observation run of the LIGO/Virgo gravitational-wave observatories
    have increased the number of GW observations to around 90
    \cite{GWTC-3}.
These new observations extend to significantly higher mass
    \cite{LIGO-O3-GW190521-discovery,LIGO-O3-GW190521-implications},
    within the lower mass gap \cite{GW190814},
    more extreme mass ratios \cite{GW190412, GW190814},
    and multiple \nsbh{} events \cite{LIGO-O3-O3b-NSBH,GWTC-2, GWTC-2p1, GWTC-3}.
These objects' existence strongly constrain formation models.
For example, the surprising 
    secondary
    mass of GW190814 \cite{GW190814} may constrain
    models for SN, as not all such models can produce events in the lower mass gap
    \cite{popsyn-nsbh-Pawel-EjectaStudy2020}.
As we show below, a joint analysis of the whole population provides very stringent
    model constraints, which are difficult to satisfy without
    allowing for the possibility of substantial systematic error.

In the analysis below, we use Gaussian approximations to the likelihood of the
    confident events (excluding GW190521) from the third observing run of 
    LIGO/Virgo \cite{LIGO-O3-O3b-RP},
    as a function of mass alone.
GW190521 in particular is excluded as it would require non-standard physical
    assumptions to be formed in isolated binary evolution
    \cite{BelczynskiGW190521}
    that we do not test in this current study.
Alternatively, it may have formed through dynamical interactions
    in a dense stellar cluster
\cite{Gayathri2022, Gamba2021}.
In our preliminary analysis, we marginalize over and do not attempt to reproduce binary spin.
The specific parameters of these single-event likelihoods are described in \cite{nal-methods-paper}.

\end{subsubsection}

\end{subsection}
 
\section{One-dimensional study: Natal kick velocity}
\label{sec:kick}
\label{chap:kick}

To illustrate our methods,
    we first use  a simple low-dimensional approach that employs 
    a series of \startrack{} models in which we alter only one
    evolutionary parameter:
    the 
    dispersion parameter of the NS/BH \kick velocity
    ($\STkick$).
This parameter has been
    thoroughly explored in previous work,
    notably in Wysocki et al \cite{Wysocki2018}.
We emphasize that in previous work with \startrack{},
    two types of models have been examined:
    models with \emph{fallback-suppressed} \kicks{},
    where the kick distribution depends on the amount of fallback;
    and homogeneous \kicks{}, where the same distribution
    is applied to all compact objects, independent of fallback.
In this work, we explore the latter.

As we show below,
    we find that for these models the merger rate varies strongly and the mass
    distribution weakly versus \kick{} strength.
    The merger rate is highly informative and the mass distribution weakly informative about 
    \kicks{}.

\begin{figure*}
\includegraphics[scale=0.5]{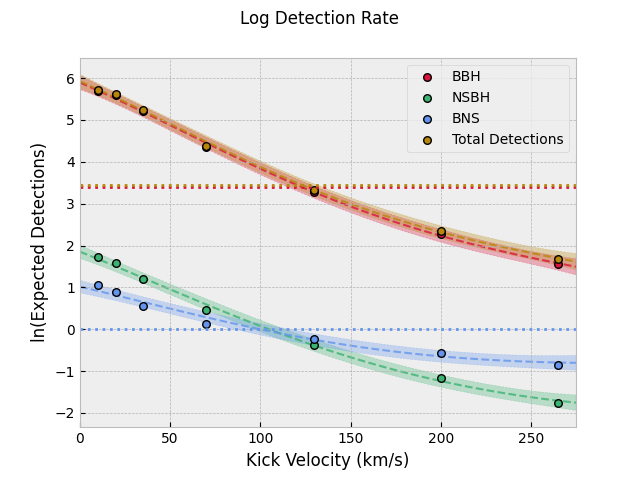}
\includegraphics[scale=0.5]{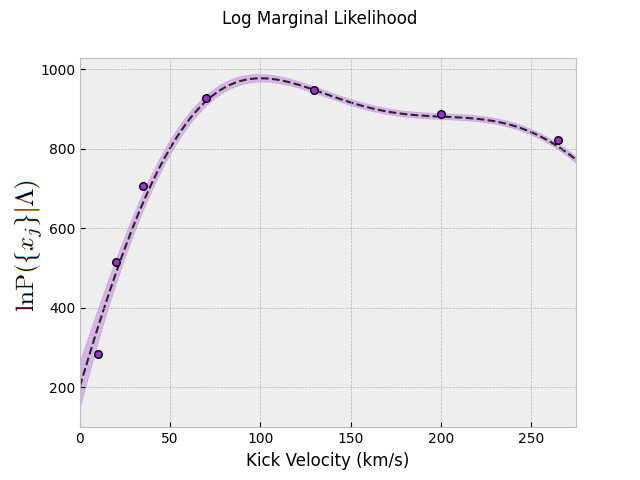}
\caption{ \label{fig:Kick:Comparison} \textbf{Likelihood versus kick in O3a}: 
\emph{Left}:
    The log of the expected number of detections 
        for models with a particular kick velocity.
    The scattered points are values calculated by \startrack{} models.
    The dashed line is the mean prediction of the Gaussian Process interpolator,
        where the shaded regions are regions of $\STkick$.
    The horizontal dotted lines are the number of GW observations in O3a.
\emph{Right}:
    The joint likelihood distribution of the kick parameter,
        considering the merger rate
        as well as the distribution of observations in O3a.
    Again, the scattered points represent \startrack{} Models,
        the dashed line represents the mean prediction,
        and the region of $\STkick$ is shaded.
    The interpolation of the joint likelihood follows the second method
        outlined in Section \ref{sec:interpolation},
        incorporating information from each type of event rate
        and the shape likelihood.
}
\end{figure*}

\begin{figure}
    \begin{centering}
    \includegraphics[scale=0.5]{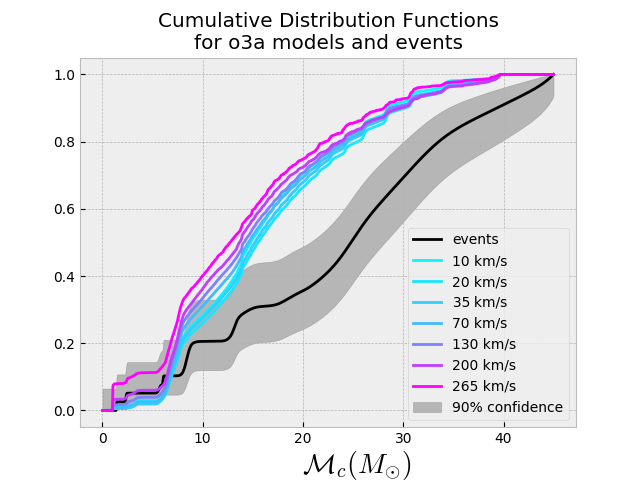}
    \caption{
        \label{fig:kick_cdf}
        \textbf{Chirp mass cumulative distributions: One-dimensional \kick{} velocity survey}:
        Cumulative Distribution Function (CDF) in $\mc$ for models M13-19,
        for which $\STkick$ is varied.
     For comparison,
        a solid black line indicates the CDF for observations in the first part
        of LIGO/Virgo's third observing run
        (shaded gray region indicates 90\% symmetric credible interval for
        those observations).
        }
    \end{centering}
\end{figure}

For these models,
    as the \kick{} velocity increases,
    an increasing fraction of compact binaries are disrupted,
    causing the compact object merger and detection rate
    \emph{overall} to vary strongly with kick strength.
For example,
    Figure  \ref{fig:Kick:Comparison}
    shows the predicted detections for an observing run as a function of
    \kick{} strength, compared to the first part of LIGO's third observing run
    (O3a) \cite{GWTC-2, GWTC-2p1}.
Given the other assumptions in our binary evolution simulations
    (refer to Section \ref{sec:models}),
    the strength of fallback-independent \kicks{} would
    need to be large (between $70\unit{km/s}$ and $200 \unit{km/s}$)
    to explain the overall number of observations in O3a.

By contrast,
    Figure \ref{fig:kick_cdf} shows the chirp mass distribution
    of observable gravitational-wave events for several choices of kick strength.
Though the overall number varies substantially,
    the \emph{shape} of the mass distribution changes relatively little,
    except for the relative proportions of \bns{} and \nsbh{} binaries
    (i.e., the shape of the chirp mass distribution at very low mass).

In this case,
    we gain relatively little additional information from comparing
    the observed masses to this limited model family
    (keeping in mind that the M13-19 models do not include
    updated supernova physics such as the delayed timescale of
    shock propagation which is necessary for the prediction
    of events in the lower mass gap such as
    GW190814 \cite{GW190814};
    see Figure \ref{fig:M15}).
The mismatch between the shape of the observed population and the simulated
    population for this one-dimensional study (see Figure \ref{fig:kick_cdf}),
    regardless of the chosen $\STkick$,
    suggests that
    varying assumptions beyond just the dispersion of \kicks 
    is necessary to develop our understanding of binary evolution.
The higher-dimensional study better describes the properties
    of the observed detections.

It follows that in order to properly constrain the shape of the
    population,
a higher-dimensional study has the capacity to further
    constrain these models in a way that this one-dimensional
    studies does not.

Figure \ref{fig:Kick:Comparison} shows the key result of this one-dimensional
    study:
    the marginal likelihood of each binary evolution model as a
    function of kick velocity.
This likelihood incorporates not only information about the detection rate,
    but the properties of the observed gravitational-wave population
    (refer to Section \ref{sec:inference} for details about this calculation).

This likelihood can be used in a straightforward way to generate a posterior,
    by assuming a uniform prior in binary evolution parameters;
    we do not demonstrate this for the one-dimensional study,
    but provide examples in the four-dimensional study in
    the next section.
As described above, 
    our calculations favor 
    \kickAdj{} \kicks{}
     ($>70\unit{km/s}$)
    to explain the observed merger rates and masses.
For O3 and this limited parameter survey,
    merger rates for all three event classes happen to be consistent with the
    same kick velocity: $\simeq 125 \unit{km/s}$ 
    (model M15; see Figure \ref{fig:M15} for the distribution of predicted
    detections for this model).

As discussed previously,
    systematic uncertainties
    (i.e., we only vary a subset of all binary evolution parameters,
    and the neglected parameters have a non-negligible effect)
    and uncertainties 
    in our ability to correctly quantify GW survey
    systematics with our simple approximations propagate
    into uncertainties in our marginal likelihoods.
We do note, however, that less information is lost due to our
    likelihood model than waveform systematics
    \cite{nal-methods-paper}.
For simplicity, however,
    we generate posterior distributions for our
    binary evolution parameters (here, \kick{} $\STkick$) without
    propagating these systematic effects
    (This choice also doesn't over-smooth incorrectly,
    inappropriate for the net correlated impact of some
    systematics like input SFR normalization,
    which influence all models with a common multiplicative factor.)

\section{Four-dimensional study}
\label{sec:results-full}
\begin{table*}[!ht]
\centering
\begin{tabular}{|c|c|c|c|c|}
\hline
Model IDs & Successful Models & Sampling & Best ID & Best lnL \\
\hline
K0100-K0399 & 284 & Uniform $\STFa \in [0.1,1.0]$, $\STBeta \in [0.,1.]$, $\STkick \in [0., 265.]$, $\STWindOne \in [0.2, 1.0]$ & K0358 &55.468\\
\hline
K0400-K0499 & 97 & Uniform $\STFa \in [0.5,1.0]$, $\STBeta \in [0.,1.]$, $\STkick \in [20., 150.]$, $\STWindOne \in [0.2, 1.0]$ & K0483 &62.001\\
\hline
K0500-K0519 & 20 & Samples from truncated Gaussian fit to K0100-K0499 & K0506 & 67.117\\
\hline
K0520-K0559 & 40 & Samples from truncated Gaussian fit to K0100-K0519 & K0559 & 67.995\\
\hline
K0560-K0563 & 4 & Cherry-picked & K0563 & 65.437 \\
\hline
\end{tabular}
\caption{\label{tab:K_parameter_sampling}
\textbf{Sampling methods for four dimensional model space:}
Each row depicts the sampling method used for each subset of the four-dimensional 
    parameter space models.
With each subset of models, our inferred joint likelihood distribution
    is better defined and understood through iterative improvements.
Truncated Gaussians rely on GWALK \cite{nal-methods-paper}.
For all simulation parameters, see Appendix \ref{ap:tables}.
}
\end{table*}

\begin{figure}
\centering
\includegraphics[width=3.375 in]{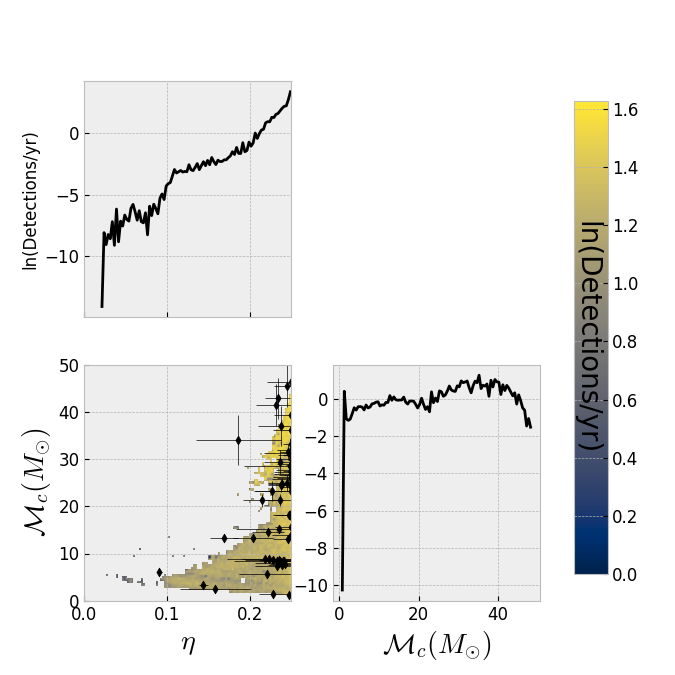}
\caption{\label{fig:Kbest}
Compare Figure \ref{fig:M15}.
\emph{Off-diagonal:}
A two-dimensional density of predicted detections
    for the \Kbest{} model (in $\mc$ and $\eta$).
Also plotted are 
Maximum Likelihood Estimate (MLE)
    parameter values for confident observations
    \cite{LIGO-O3-O3b-RP}.
\emph{Diagonal:} one-dimensional detection-weighted densities
    for \Kbest{} in $\mathcal{M}_c$ and $\eta$.
For this model, we note the absence of a distinct lower mass gap
compared to the M15 model presented in Figure \ref{fig:M15}
    and an upper-range of \bbh{} 
chirp
    mass which is 
    well into the 40s of solar mass.
}
\end{figure}

\begin{figure}
\includegraphics[width=\columnwidth]{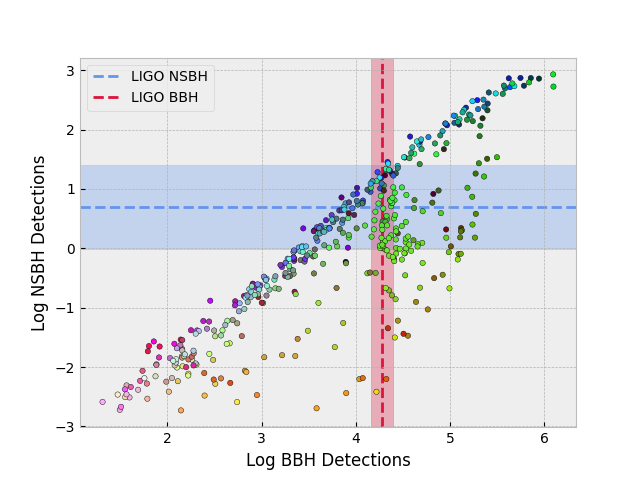}
\includegraphics[width=\columnwidth]{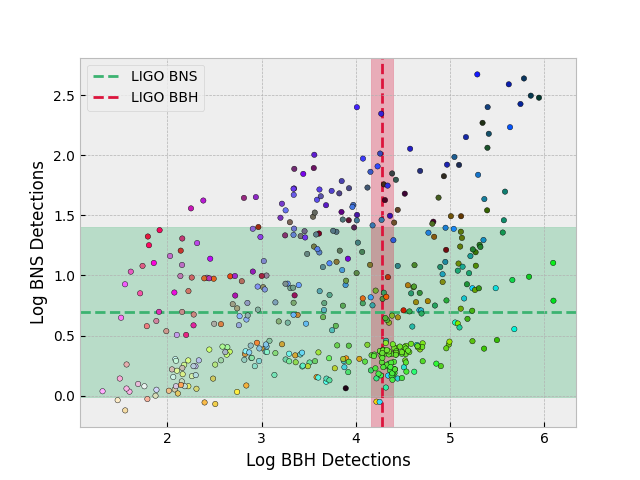}
\caption{\label{fig:Overall:RateScatter} \textbf{Expected Detection counts and empirical constraints}
Scatter plot of the number of predicted \bbh{}, \nsbh{}, and \bns{}
    detections for the
    four-dimensional simulated model family examined in this work,
    in comparison with the observed number seen in the first three observing runs
    of ground-based gravitational-wave interferometers
    (consistent with \cite{LIGO-O3-O3b-RP}).
Detection rates are
    expressed in natural logarithmic scale.
Dashed lines indicate observations;
    with shaded regions indicating counting error 
$1/\sqrt{n_{obs}}$;
    compare to \cite{Belczynski2020-EvolutionaryRoads}.
The dots with dark borders denote results from our simulations.
$\STkick \in [0.,265.]$ km/s, $\STFa \in [0.1,1.0]$, and $\STWindOne \in [0.2, 1.0]$
    encode the red, green, and blue pixel data ($\in [0,1]$) for each model.
This color map aids in identifying ranges of formation parameter values
    in the limited dimensions of this figure.
The cluster of green samples largely correspond to simulations proposed
    using the truncated Gaussian model
    (see Table \ref{tab:K_parameter_sampling}).
}
\end{figure}
We now apply our method to the four-dimensional model family performed for this work.
    These \NsimFull simulations cover three salient binary evolution parameters:
    $\STkick$, the 
    dispersion parameter for the Maxwellian distribution
    from which \kick velocities are drawn.
    (a single Maxwellian is used for BH and NS kicks, which are not reduced by fallback);
    $\STFa$, a parameter characterizing the efficiency of mass transfer;
    and $\beta$, the specific angular momentum of the ejected material.  
Unlike the simulations described earlier,
    we also account for the suppression of mass-loss due to stellar winds
    described by \cite{Belczynski2020b};
    we vary the parameter $\STWindOne$ to account for this suppression
    in hydrogen-dominated stars,
    but keep $\STWindTwo = 1$ fixed -- indicating that mass-loss is not suppressed
    for helium-dominated stars.
Belczynski et al. \cite{Belczynski2020b} have demonstrated that
    a sufficient reduction ($\STWindOne \approx 0.2$)
    allows for the formation of much larger remnant objects
    from isolated binary evolution (such as GW190521).
We begin with a uniform sampling in the space of these four formation parameters,
    and iteratively refine a truncated Gaussian model to sample from regions
    of higher likelihood.
Table \ref{tab:K_parameter_sampling} highlights that though the space 
    of formation parameters is first explored through samples drawn uniformly
    in the parameter space,
    at each iteration we were able to refine our model,
    and starting with the models labeled K0400 and higher,
    samples are drawn from a four-dimensional truncated Gaussian
    fit to the peak likelihood at hand-picked intervals
    (where points sampled from the Gaussian
    outside the bounds of the space of our formation parameters
    are disregarded).
This allows us to iteratively refine the peak in joint likelihood.

Furthermore, for this family of models,
    we adopted a SN engine with \engine{} shock propagation,
    as needed to reproduce objects in the lower mass gap like the secondary
    in GW190814 
    \cite{sn-RemnantMasses-Fryer2011,popsyn-nsbh-Pawel-EjectaStudy2020}.
We also consider the effect of weak pair-production instability
    in order to model pair instability and pulsational pair instability supernova;
    the latter of which expand the range of predicted high-mass remnant objects
    \cite{Belczynski2016PSN,Belczynski2020-EvolutionaryRoads}.
The simulations cover a broad range of possibilities,
    including models which are consistent with most
    of the confident detections reported in O3;
    Figure \ref{fig:Kbest} further illustrates this point
    \cite{LIGO-O3-O3b-RP}.

\begin{figure}
\includegraphics[width=3.375 in]{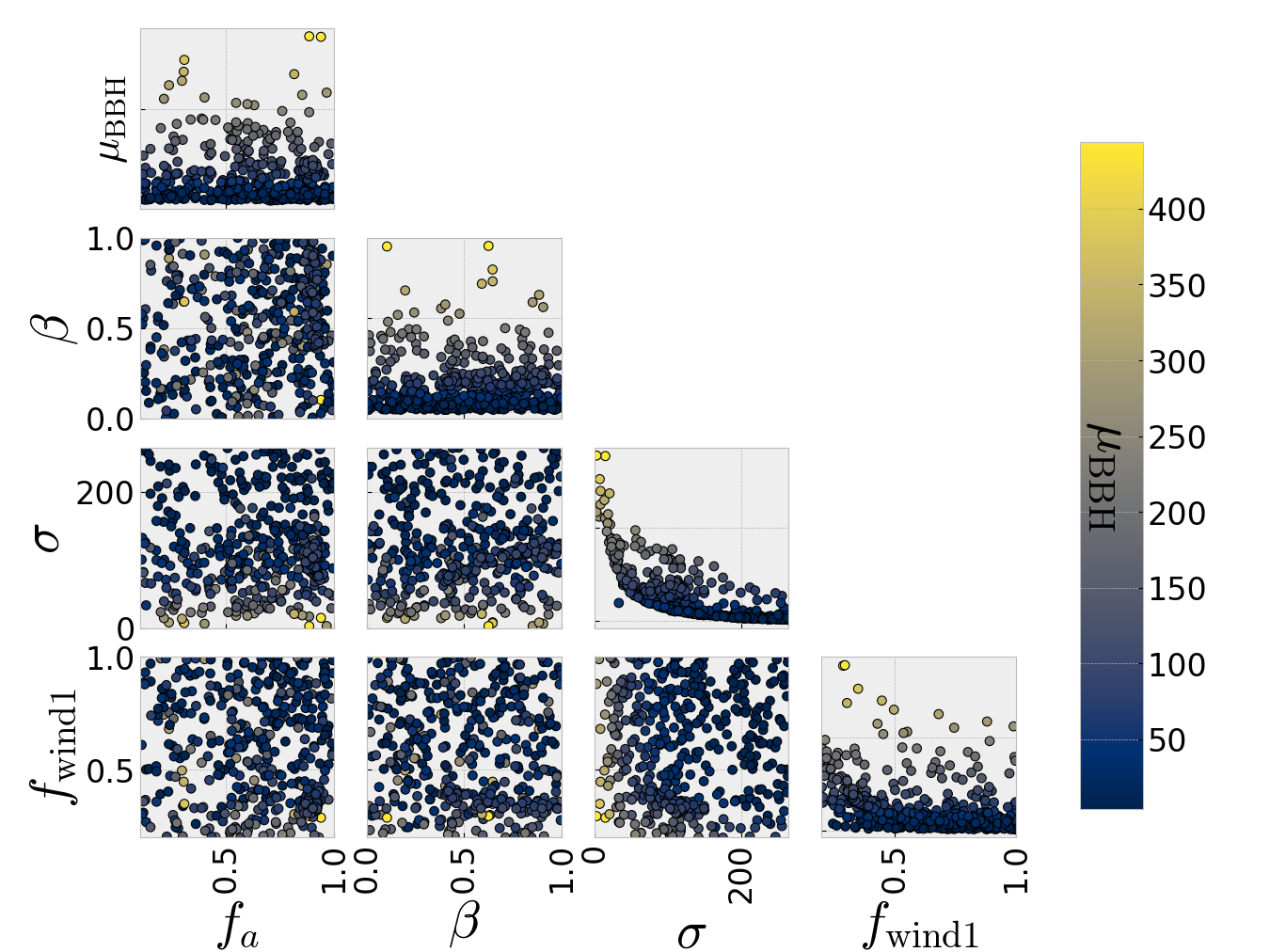}
\includegraphics[width=3.375 in]{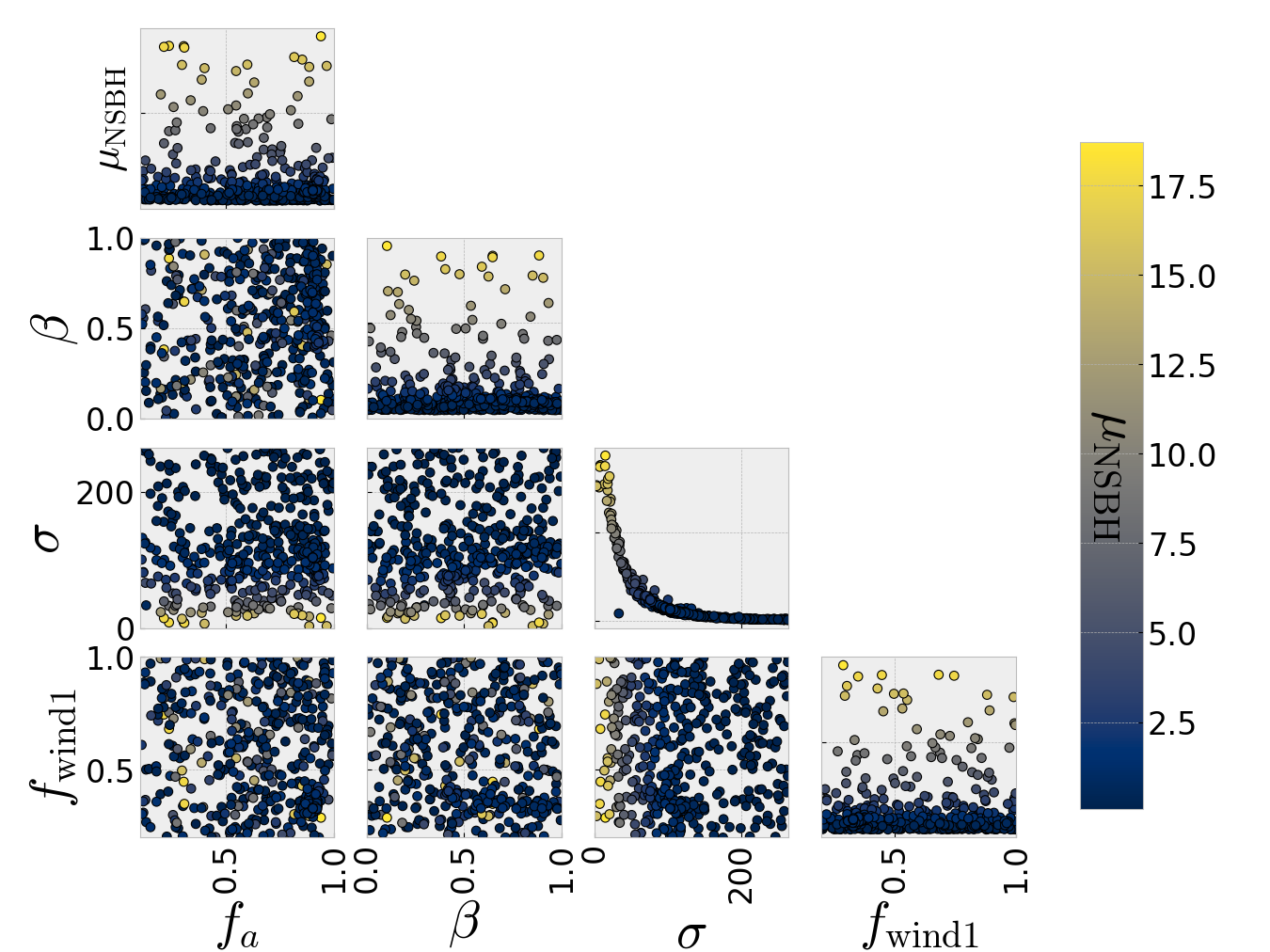}
\includegraphics[width=3.375 in]{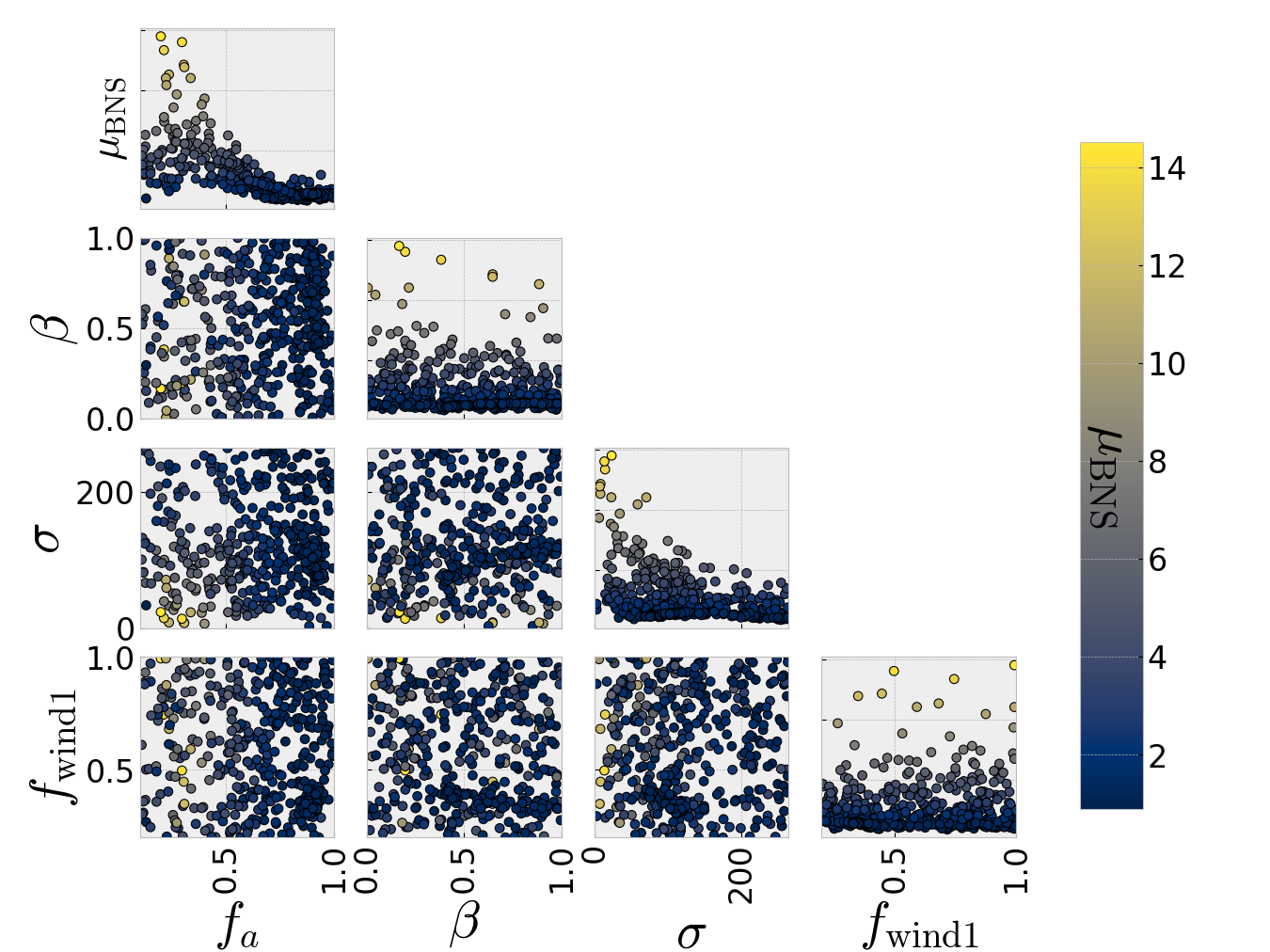}
\caption{\label{fig:DetectionRateVersusParameters} \textbf{Detection rates versus formation parameters}:
Scatter-plot in formation parameter space of
    \bbh{} (top), \nsbh{} (center), and \bns{} (bottom) mergers,
    where the color scale indicates the detection rate for each simulation.
Off-diagonal plots show a two-dimensional scatter in formation parameters
    ($\STFa$, $\STBeta$, $\STkick$, and $\STWindOne$),
    while diagonal plots show a one-dimensional scatter
    (where the y-axis is the detection rate; as also indicated by the color).
Detection rates are estimated with the two-detector sensitivity and
    observing time from O3 \cite{GWTC-2, GWTC-3}.
}
\end{figure}

\begin{figure}
\includegraphics[width=\columnwidth]{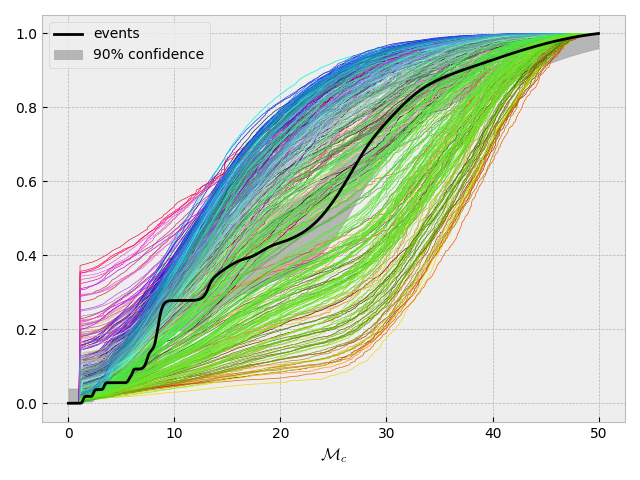}
\caption{\textbf{Chirp mass cumulative distributions: Four-dimensional model survey}:
\label{fig:Overall:CDFScatter}
Chirp mass CDFs for the four-dimensional model family described in 
Section \ref{sec:results-full}.
    The color scale encodes three model parameters as RGB values
    (RGB encode $\STkick$, $\STFa$, $\STWindOne$ respectively).
For comparison,
    the solid black line and gray region show the cumulative distribution function for
    O3 observations and the corresponding 90\% credible interval, respectively.
}
\end{figure}

To highlight the diversity of these simulations,
    Figure \ref{fig:Overall:CDFScatter} shows the chirp mass
    distributions for our simulations.
Similarly,  Figure \ref{fig:Overall:RateScatter} shows the
    expected number of \bbh{} and \bns{} observations in O3,
    compared to current population models.
Figure \ref{fig:DetectionRateVersusParameters}
    catches a further glimpse into the dependence of the detection rate
    on each formation parameter and combination thereof.

Different formation parameters have strong impacts on different populations.
For example,
    Figure \ref{fig:DetectionRateVersusParameters}
    shows how the detection rate changes versus the parameters in our study.
For \bbh{}, the merger rate is principally determined by $\STkick$,
    with a sub-dominant impact from $\STWindOne$ and other parameters.
For \bns{}, the merger rate is determined by 
    both $\STFa$ and $\STkick$.
For \nsbh{}, the merger rate is principally determined by 
    $\STkick$, in a very tightly dependent manner.

Figure \ref{fig:DetectionRateVersusParameters}
    also visually suggests what combinations of parameters are required
    to reproduce current event counts.
For these models, a \kickAdj{} $\STkick$ is required to predict the correct
    amount of \bbh{} and \nsbh{} detections.
With only two reported observations so far,
    the \bns{} and \nsbh{} merger rates are highly uncertain,
    and the observed counts are consistent with
    (but in modest tension with)
    what's expected for 
\kickAdj{} \kicks{}.

Note that while the reported number of \bbh{} observations seems very strongly constraining,
    systematic uncertainties highlighted
    in the previous section associated with sub-dominant parameters and
    input uncertainties imply that the absolute merger rate must be interpreted with caution.
Similarly, even adopting the \engine{} SN engine and
    even not aggressively adjusting physics associated with the pair
    instability gap,
    the reported chirp mass distribution can encompass most of the observations reported to date
    (see Figure \ref{fig:Overall:CDFScatter}).

\begin{figure}
\centering
\includegraphics[width=3.375 in]{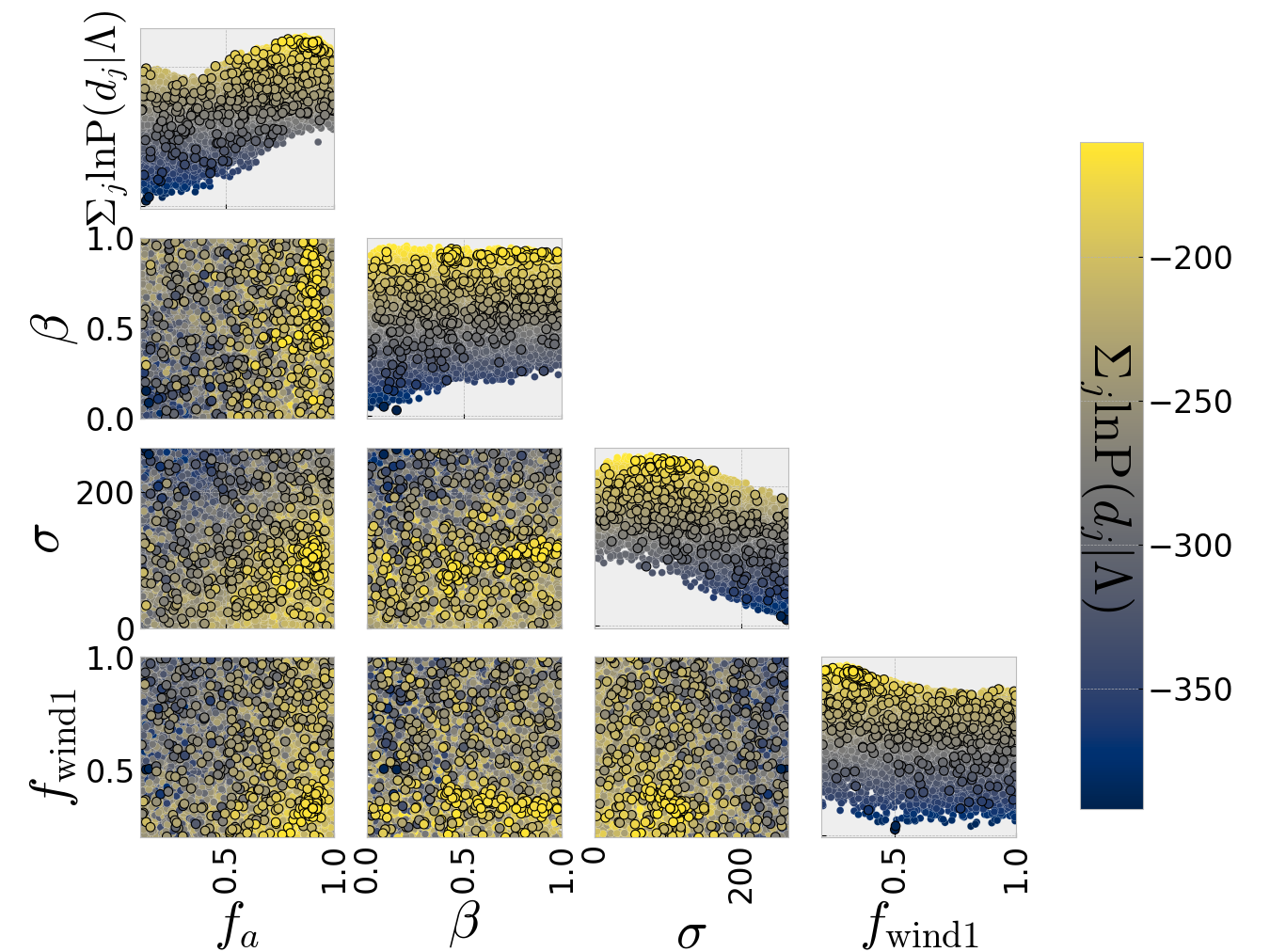}
\includegraphics[width=3.375 in]{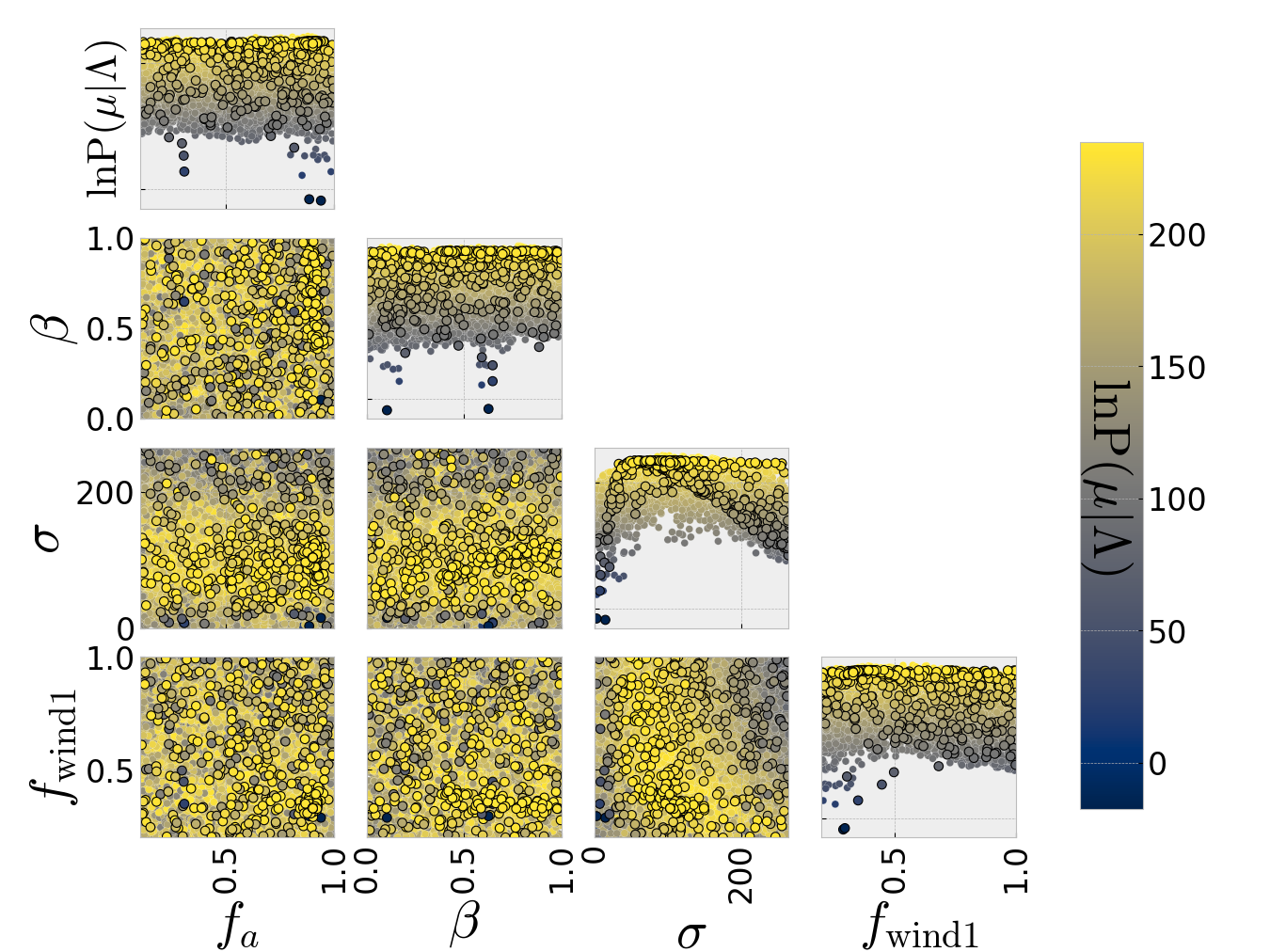}
\includegraphics[width=3.375 in]{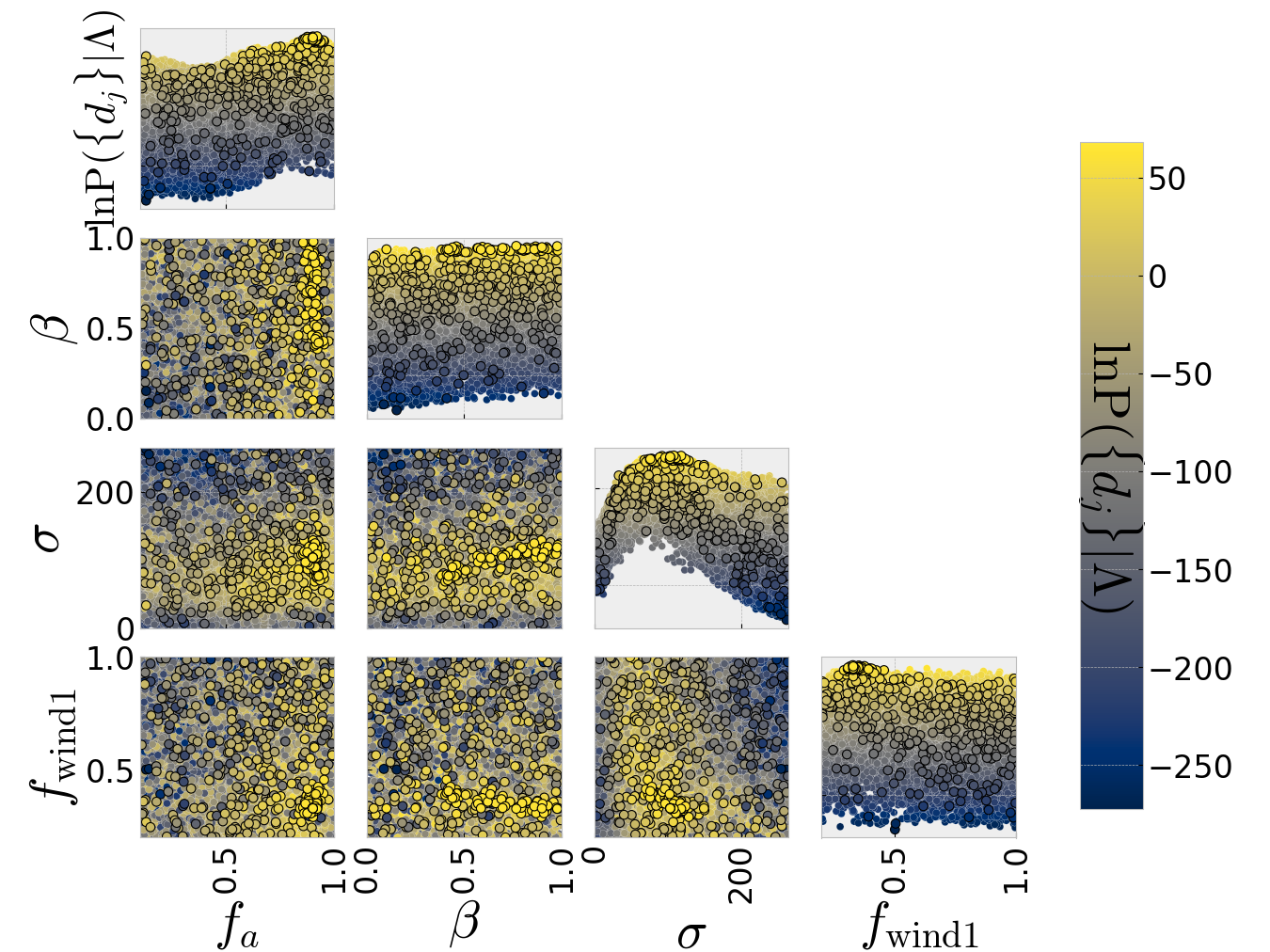}
\caption{ \label{fig:full-parameter-likelihood} \textbf{Inferences on four-dimensional model constraints: }: 
Scatter-plot in formation parameter space for quantities of interest
    (indicated by color):
    \emph{shape likelihood} (top), \emph{rate likelihood} (center), and \emph{joint likelihood} (bottom).
Off-diagonal plots show a two-dimensional scatter in formation parameters
    ($\STFa$, $\STBeta$, $\STkick$, and $\STWindOne$),
    while diagonal plots show a one-dimensional scatter
    (where the y-axis is the quantity of interest; as also indicated by the color).
Larger dots with dark borders denote results from our full simulations;
Smaller background dots with white borders are posterior samples drawn from a uniform prior,
    weighted by interpolated likelihood.
    The scale indicated for the rate and shape likelihood can be used to
        asses their relative importance, as the joint likelihood is
        the sum (in natural log) of the two quantities.
    In contrast to Figure \ref{fig:Kick:Comparison},
        the joint likelihood is interpolated directly
        and sampled according to the method outlined in Section
        \ref{sec:interpolation}.
}
\end{figure}

We can learn as much about our formation parameters from the shape of the
    mass distribution of predicted mergers as we can from the detection rates.
We have evaluated the agreement of the masses of each sample detection from
    each simulation with each confident gravitational-wave
    observation from the first three observing runs of ground-based
    detectors (consistent with \cite{LIGO-O3-O3b-RP}).
This information about the shape of the distribution is 
    described by the \emph{shape likelihood} 
    (see Section \ref{sec:shape-likelihood}).
The full inhomogeneous Poisson likelihood can be re-composed
    from the component \emph{rate} and \emph{shape likelihood} components,
    and we refer to this as the \emph{joint likelihood}
    (see Section \ref{sec:shape-likelihood}).
Figure \ref{fig:full-parameter-likelihood} displays
    the rate, shape, and joint likelihood to a constant
    (as a function of formation parameters).
    For these models, both the rate and shape likelihood
    carry a meaningful weight to the joint likelihood.
    This is especially noteworthy in the two-dimensional
    marginal likelihood between $\STkick$ and $\STWindOne$.

Figure \ref{fig:full-parameter-likelihood} indicates 
    that the shape of the distribution favors
    high accretion of mass lost by a donor during 
    Roche-lobe mass transfer ($\STFa$)
    as well as moderate specific angular momentum of ejected material
    ($\STBeta$).
Furthermore, there is a preference for moderate \kicks{}.
We also find that reduced mass loss rates from stellar wind
    are substantially favored (indicated by a low $\STWindOne$),
    consistent with \cite{Belczynski2020-EvolutionaryRoads}.
These trends are roughly followed by the joint likelihood as well.

\begin{figure*}
\centering
\includegraphics[width=6.750 in]{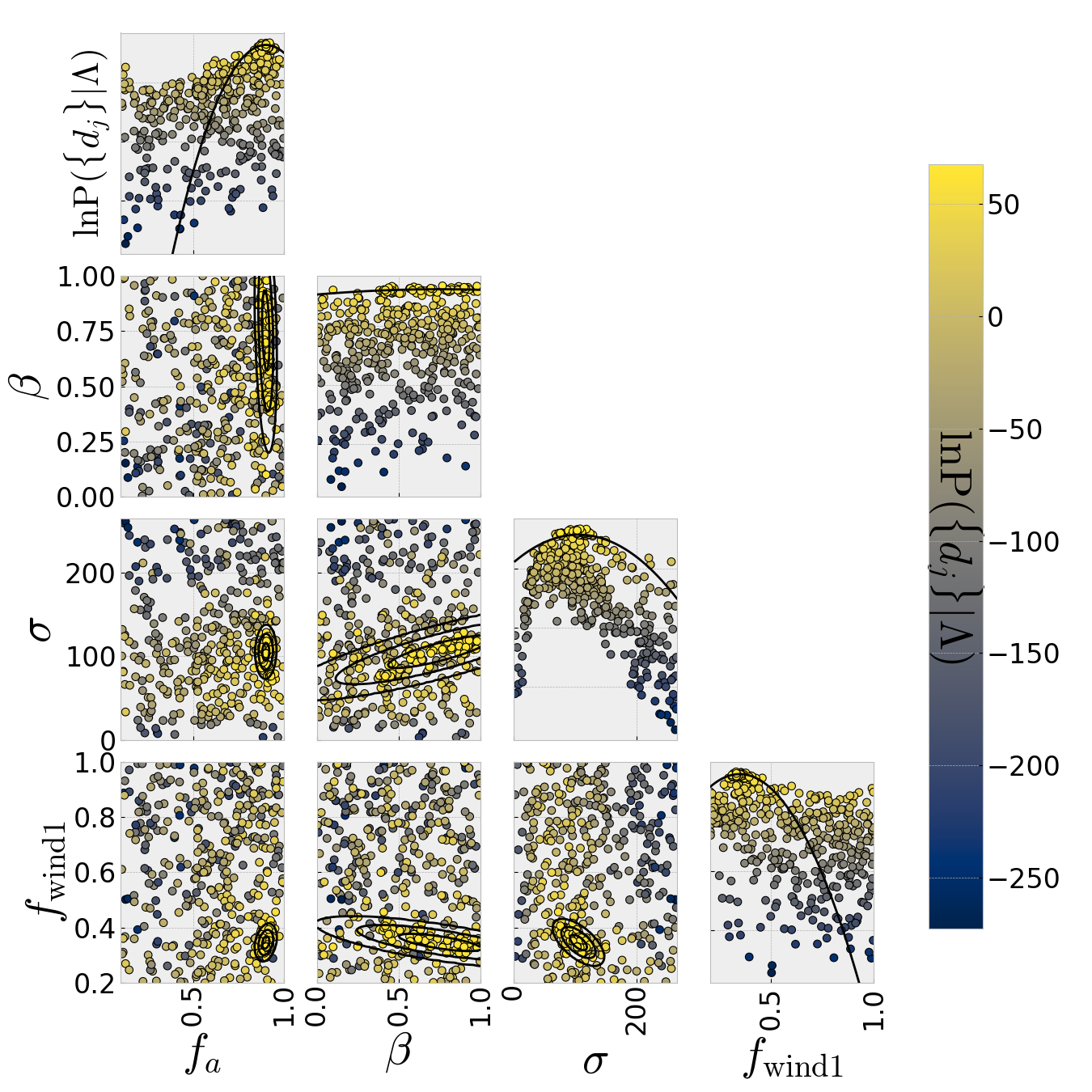}
\caption{\label{fig:formation-gaussian} \textbf{Approximate Gaussian Posterior}
Large scatter points with dark borders represent each simulated universe
    (as in Figure \ref{fig:full-parameter-likelihood}).
A Gaussian approximation to the posterior has been drawn in black
    on one-dimensional diagonal plots;
    contours enclosing sampled regions of 1, 2, and 3
    standard deviations are drawn on two-dimensional plots. 
Keep in mind that a Gaussian appears quadratic in the log space.
This final approximation was fit to simulations within 7
    of the maximum $\mathrm{ln}\mathrm{P}(\{d_j\} | \Lambda)$ (27 simulations).
Samples are drawn from Gaussians to propose the next batch of simulations,
    in a way that iteratively refines our ability to describe
    the relationships between parameters (see Table \ref{tab:K_parameter_sampling}).
See also Appendix \ref{ap:FormationGaussian} for the properties
    of this Gaussian.
}
\end{figure*}

For the most general likelihood distribution without a single
    clear peak,
    the interpolation of the rate, shape, and joint likelihood
    (as in Figure \ref{fig:full-parameter-likelihood})
    best characterizes what we have learned about our formation parameter
    assumptions by comparing simulations to observations.
These interpolated likelihoods can then describe a posterior
    in those formation parameters in the region of highest likelihood.
We don't claim to do this, as such an approach overstates our confidence
    in model systematics:
    we've only explored a small number of the many important parameters
    that impact binary evolution,
    and many omitted parameters and physics are well-known to have a 
    significant impact on observables. 

This interpolated likelihood can then be sampled from,
    in order to propose new simulations and
    further grow our knowledge about the space of our
    assumed formation parameters.
However, we have found that in this specific case
    there is a peak in likelihood that can be 
    described locally by a Gaussian
    (see Figure \ref{fig:formation-gaussian}).
This normal approximation allows us to characterize the best fitting parameters
    and their correlations.Fitting a bounded multivariate normal distribution
    to this likelihood greatly simplifies the process of sampling from it,
    and therefore we use this approximation to the likelihood in the region
    around this local maximum in order to propose new simulations.
This sampling process only requires using known methods
    \cite{2020SciPy-NMeth}
    of sampling from a multivariate normal distribution
    and discarding points sampled outside of the boundary
    (the limits of our formation parameter space)
    in order to propose new simulations.

In the future, this sampling may be done automatically by an algorithm.
At present, each iteration requires choices about how sampling is
    accomplished.
Table \ref{tab:K_parameter_sampling} describes this process
    for the K series of models presented here.

Through interpolating and/or modeling this joint likelihood
    and drawing samples from a likelihood model
    before launching subsequent simulations,
    we iteratively improve our understanding of the relationship
    between model parameters and likelihood in the four-dimensional space
    (see Table \ref{tab:K_parameter_sampling}).
This method of hierarchical inference for constraining
    the isolated binary evolution formation channel
    for gravitational-wave populations
    is a benchmark for future studies of higher-dimensional
    population models.

\begin{table*}[!ht]
\centering
\begin{tabular}{|c|c|c|c|c|c|c|c|c|c|c|c|}
\hline
model &  $\STFa$ &  $\STBeta$ &  $\STkick$ &  $\STWindOne$ &  $\STWindTwo$ &  $\mu_{\mathrm{\bbh{}}}$ &  $\mu_{\mathrm{\bns{}}}$ &  $\mu_{\mathrm{\nsbh{}}}$ &  lnL\_rate &  lnL\_shape &  lnL\_joint \\
\hline
K0559 &  0.922 &  0.768 &  108.299 &  0.328 &  1.000 &  95.015 &  1.453 &  1.218 &  231.346 &  -163.351 &  67.995 \\
\hline
K0556 &  0.905 &  0.979 &  118.317 &  0.328 &  1.000 &  88.926 &  1.435 &  0.827 &  232.277 &  -164.747 &  67.529 \\
\hline
K0506 &  0.898 &  0.459 &  92.364 &  0.353 &  1.000 &  92.756 &  1.317 &  1.557 &  231.964 &  -164.848 &  67.117 \\
\hline
K0524 &  0.876 &  0.867 &  103.613 &  0.352 &  1.000 &  79.303 &  1.414 &  1.243 &  234.044 &  -167.049 &  66.995 \\
\hline
K0542 &  0.907 &  0.594 &  101.609 &  0.376 &  1.000 &  72.266 &  1.369 &  1.365 &  234.436 &  -168.259 &  66.177 \\
\hline
\end{tabular}
 \caption{\label{tab:Kbest}
The parameters associated with the five best models,
    ranked by the inhomogeneous Poisson likelihood
    (see Section \ref{sec:inference}).
}
\end{table*}

We have evaluated the marginal likelihood for each of our simulations,
    with the raw data available in an associated data release.
Table \ref{tab:Kbest} shows the parameters of the five most highly ranked simulations,
    and the remainder are included in Appendix \ref{ap:tables}.

\section{Conclusions}
\label{sec:conclusions}
In this work we have demonstrated a consolidated and efficient way to perform
    Bayesian inference on sparsely-sampled simulations of compact binary
    formation with isolated binary evolution,
    where each simulation provides only a weighted sample of events.
We applied our technique to compare a small collection of \startrack{} binary
    evolution simulations to the compact binary population reported by
    the LIGO/Virgo collaboration
    \cite{GWTC-1, GWTC-2, GWTC-2p1, GWTC-2, LIGO-O3-O3b-RP}.
Consistent with prior work,
    and with observations of galactic X-ray binaries proper motions
    (see, e.g. Table 7 in \cite{Belczynski2016}),
    we conclude that \kicks{} would need to be modest but nonzero
    to explain the numbers and properties of observed binaries.
Previous proof-of-concept studies
    \cite{ROSPSmoreconstraints, Wysocki2018}
    have demonstrated the merger rate and mass distribution to be very
    effective at discriminating between 
    different evolutionary
    models within isolated binary evolution.
With the increasing number of observed gravitational-wave events
    as ground-based observatories like LIGO, Virgo, and KAGRA
    continue to scan the universe for merging compact binaries,
    we anticipate 
    this method can enable tight constraints on physical processes
    within isolated binary evolution
    \cite{fritschel2020instrument,Virgo,CalibrationPaper,P1200087,O3-Detector,Kagra}.

Other groups (including \cite{broekgaarden2021formation, broekgaarden2021impact})
    have studied the significance of the two \nsbh{} events
    observed in the third observing run of the modern ground-based
    gravitational-wave observatories \cite{LIGO-O3-O3b-NSBH}.
We concur with their findings that these observations are essential
    to constraining the formation of compact objects
    through isolated binary evolution,
    and we demonstrate this by considering the strong
    correlation between \kicks{} and \nsbh detection rates
    (see Figure \ref{fig:DetectionRateVersusParameters}).

Other groups have also developed methods to infer what kinds of binary evolution models are compatible with
observations.
For example, one investigation     \cite{COSMICWong2022} involved backwards-propagating binaries from their final state
to consistent progenitor configurations,  recovering ranges for both  physical (binary mass and orbit) and binary evolution model
parameters.   
Meanwhile, another group \cite{Taylor2018popsyn}
    has demonstrated a method for constraining individual
    assumptions concurrently through iteratively refining 
    a one-dimensional prior for each continuous parameter.

In these and other studies, previous works  have highlighted the challenge in thoroughly investigating the many  uncertain assumptions
    implicit in current binary evolution models; see, e.g., the discussion in  \cite{COSMICWong2022}.
That said, 
by sampling iteratively from an evolving higher-dimensional posterior,
    we demonstrate that constraints can be drawn on many of
    the continuously variable parameters 
    (as opposed to assumptions with only discrete settings) at once.
Doing so allows us to explore correlations and confounding effects
    between formation parameters and narrow down the space
    of our assumptions.

Various groups have carried out the integration of marginalized likelihoods
    on binary evolution simulations in the past;
    their methods included drawing samples from the posterior
    for each event \citeMCMCpopsyn,
    kernel-density estimates \citeKDEpopsyn,
    and Gaussian Mixture Models (GMM) \citeGMMpopsyn.
Our method for evaluating these marginalized likelihoods
    offers several technical advantages;
    we evaluate a well-constrained likelihood model
    \cite{nal-chieff-paper, nal-methods-paper} for each single-event likelihood,
    allowing us to integrate over the merger population directly.
We can therefore more confidently assess ``outlier'' events and otherwise
    make best use of all available simulation data.
We can also immediately work with all simulation observables,
    without worrying about artificially introducing features with
    a smoothing algorithm.
This approach can be applied immediately to other compact binary formation models
    of comparable sophistication,
    including models for isolated binary evolution
    \cite{2020arXiv200903911S, 2020MNRAS.tmp.3003D},
    dense interacting stellar clusters
    \cite{2019PhRvD.100d3027R,2020arXiv200610771K},
    and active galactic nuclei
    \cite{gwastro-agndisk-McKernanPredictMassSpin2019}.

Our approach contrasts with the other two most widely-used
    approaches in population inference.
One long-term robust approach is to eschew a concrete model,
    constructing a de-facto non-parametric distribution for the rate density
    $\rho(\BinaryParameters)$ versus binary parameters $\BinaryParameters$
    using the observed set of events,
    the known or search measurable selection biases,
    and classical statistical techniques. 
For example, with sufficiently many events,
    a simple weighted cumulative distribution will approximate the
    underlying chirp mass distribution
\cite{DominikIII,2015MNRAS.450L..85M}.   
Alternatively, 
    ad-hoc models are also widely used to address broad questions,
    when overwhelming statistics aren't available for a non-parametric approach.
For example,
    if coalescing binary black holes form from an isolated but interacting pair of stars,
    their initial conditions and interactions could imply
    the black hole spins are tightly aligned with the orbital angular momentum
    \cite{2000ApJ...541..319K,gwastro-mergers-PNLock-Gerosa2013,
        GW150914-astro,Belczynski2016Nature};
    if they form in densely interacting environments, by contrast,
    the spins will have random orientations;
    and GW measurements will quickly distinguish between these options
    \cite{gwastro-PE-Salvo-EvidenceForAlignment-2015,gwastro-popsynVclusters-Rodriguez2016},
    as long as black hole spin magnitudes are not small as may be the case
    \cite{Belczynski2020-EvolutionaryRoads, Bavera_2020}. 
The mass distribution may have gaps and limits, 
    the underlying physics of supernova
    \cite{popsyn-constraint-StellarMassBHMassDistribution-Empirically-Farr2010}
    or pair instabilities that prevent black hole formation by very massive stars
    \cite{Belczynski2016PSN}.  
The mass and spin distributions may provide insight into the
    supernova central engine and angular momentum transport in massive stars
    \cite{2017MNRAS.467.2146K,sn-theory-OConnorOtt-2012,sn-RemnantMasses-Fryer2011,2012ApJ...757...69U,2005ApJ...626..350H}
    [modulo caveats due to significant uncertainties in massive binary stars'
    initial conditions \cite{2009Natur.461..373A,2014ApJ...785...83A},
    binary evolution 
    \citeRatesField,
    and stellar wind mass loss \cite{DominikI}].  

However, as observations become more informative,
    each piece of uncertain physics will produce correlated impacts on
    multiple properties of the observed binary population,
    which phenomenological or non-parametric approaches won't
    naturally identify as possessing a common origin.
Model-based approaches enable sharper constraints on uncertain
    astrophysics with broad observational consequences.

Given the substantial modeling uncertainties associated with compact
    object spins at formation and their alignment,
    in this work we have only compared compact binary
    \emph{mass} distributions to \startrack{} predictions.
Observations of black hole spin magnitudes and misalignments
    can 
    possibly
    differentiate between various formation channels
    and between different physical models of angular momentum 
    transport within a given formation channel
    \cite{2010CQGra..27k4007M,Wysocki2018,
        gwastro-PE-Salvo-EvidenceForAlignment-2015,
        Bavera_2020,2017MNRAS.471.2801S,Belczynski2020-EvolutionaryRoads,
        Banerjee2023}.
For the moment it seems as though LIGO/Virgo/KAGRA black holes form with low spins
    indicating efficient angular momentum transport
\cite{starev-Fuller-LowNatalSpin, Belczynski2020-EvolutionaryRoads}.
On the other hand,
    high-mass X-ray binaries may indicate that BHs
    form with large spins,
    yet these estimates are being questioned
    \cite{Belczynski2021Apples}.
Tidal spin up of stars was proposed for high-spinning LIGO/Virgo/KAGRA BHs,
    yet it is debated whether only the first-born
    BH can be spun-up
    \cite{2016MNRAS.462..844K,Bavera_2020},
    or does the process apply to both merging BHs
    \cite{Olejak2021}
Building upon this idea, other groups propose the black hole formed in a 
    binary system will always have small spin,
    but the second can have large spin due to pre-supernova spin-up
Practically speaking, however, the model space that must be explored is
    larger and the observational constraints far tighter
    when comparing both mass and spin distributions to current observations.
We defer a discussion of these degrees of freedom to future work.

Even without considering the GW190521 merger, we find a preference
    for the global reduction in mass-loss rates due to
    stellar wind for hydrogen-dominated stars
    introduced 
    to \startrack{}
    by Belczynski et al. in prior work
    \cite{Belczynski2020b}
    over non-reduced mass-loss rates
    (see Section \ref{sec:wind}).
The authors do not suggest that the reduction factor $\STWindOne$ is
    the final model for stellar wind;
    we only demonstrate that it is preferred over no reduction at all.
The weak wind phenomenon is still being understood observationally,
    and Hubble's ULLYSES program will provide spectral analysis of
    more bright stars formed at lower metallicities
    in the near future \cite{Bouret2021, HawcroftUllyses2023}.

To close, we emphasize the critical need to widely explore all relevant uncertainties.
Given the extremely tight constraints soon available,
    and past experience suggesting strong correlations between population synthesis
    model parameters,
    inferences based ona limited subspace of model parameters
    will inevitably be biased,
    potentially producing highly misleading conclusions about the most
    significant model parameters
Our own study illustrates the potential pitfalls of
    insufficiently broad exploration:
    the one-parameter survey which varied only kick velocity suggested that kick velocity
    could be well-constrained away from zero,
    but failed to reproduce the shape of the observed gravitational-wave population.
Because of the many uncertain
    parameters in all extant models for compact binary formation,
    care should be taken when drawing strong conclusions about
    what nature can permit; cf. 
    \cite{2020arXiv201103564W,2020arXiv201016333B, Belczynski2022}.

\begin{acknowledgments}
The authors thank Keith Riles, Zoheyr Doctor, Erika Holmbeck,
    and Tomasz Bulik for useful feedback.
ROS and VD are supported by NSF AST-1909534 and PHY-2012057.
ROS, VD, and AY are supported by NSF-PHY 2012057;
    ROS is also supported via NSF PHY-1912632 and AST-1909534.
VD is supported by an appointment to the NASA Postdoctoral Program at the NASA Goddard Space Flight Center administered by Oak Ridge Associated Universities under contract NPP-GSFC-NOV21-0031.
DW thanks the NSF (PHY-1912649, PHY-2207728) for support.
KB acknowledges support from the Polish National Science Center (NCN) grant
    Maestro (2018/30/A/ST9/00050). Special thanks go to tens of thousands of
    citizen-science project "Universe@home" (universeathome.pl) enthusiasts that
    help to develop StarTrack population synthesis code used in this study.
This material is based upon work supported by NSF’s LIGO Laboratory 
    which is a major facility fully funded by the
    National Science Foundation.
This research has made use of data,
    software and/or web tools obtained from the Gravitational Wave 
    Open Science Center (https://www.gw-openscience.org/ ),
    a service of LIGO Laboratory,
    the LIGO Scientific Collaboration and the Virgo Collaboration.
LIGO Laboratory and Advanced LIGO are funded by the 
    United States National Science Foundation (NSF) as
    well as the Science and Technology Facilities Council (STFC) 
    of the United Kingdom,
    the Max-Planck-Society (MPS), 
    and the State of Niedersachsen/Germany 
    for support of the construction of Advanced LIGO 
    and construction and operation of the GEO600 detector.
Additional support for Advanced LIGO was provided by the 
    Australian Research Council.
Virgo is funded through the European Gravitational Observatory (EGO),
    by the French Centre National de Recherche Scientifique (CNRS),
    the Italian Istituto Nazionale di Fisica Nucleare (INFN),
    and the Dutch Nikhef,
    with contributions by institutions from Belgium, Germany, Greece, Hungary,
    Ireland, Japan, Monaco, Poland, Portugal, Spain.
The authors are grateful for computational resources provided by the 
    LIGO Laboratory and supported by National Science Foundation Grants
    PHY-0757058 and PHY-0823459.
We acknowledge software packages used in this publication,
    including
    NUMPY \cite{harris2020array}, SCIPY \cite{2020SciPy-NMeth}, 
    MATPLOTLIB \cite{Hunter_2007}, 
    CYTHON \cite{behnel2011cython},
    ASTROPY \cite{astropy:2013,astropy:2018}, 
    and H5PY \cite{collette_python_hdf5_2014}.
This research was done using resources provided by the 
    Open Science Grid \citeOSG,
    which is supported by the National
    Science Foundation awards \#2030508 and \#1836650,
    and the U.S. Department of Energy's Office of Science. 
\end{acknowledgments}

\footnotesize\bibliography{Bibliography.bib}

\appendix

\section{SNR Interpolation}
\label{ap:SNR}

Integrating more than one million samples,
    and estimating the signal-to-noise-ratio (SNR) of each
    sample in a given \startrack{} model is computationally expensive.
In this section, I discuss a solution to this problem.
The SNR function itself uses lal to generate a psd for a particular
    observation run.
We developed a tool which interpolates the three-dimensional 
    function $\snr = \lalsnr(M_{det, A}, M_{det, B}, \lumdist(\redshift))$,
    in a two-dimensional training space,
    which is much more computationally efficient.
Here, $\lumdist(\redshift)$ is the Luminosity Distance, evaluated using astropy
    \cite{astropy:2013,astropy:2018},
    with the Planck2015 cosmology\cite{Planck2015}.

We start from the relation given by \cite{creighton2012}: 
\begin{equation}
    \label{eq:SNR_integral}
    \snr^2(\lambda) \approx \int df 
        \frac{\hat{h}(f;\lambda) \hat{h}^* (f;\lambda)}{\mathcal{S_n}(h)}
\end{equation}
which yields
\begin{equation}
    \label{eq:SNR_inv_dist}
    \snr(\mdeta, \mdetb, \lumdist(\redshift)) \approx \lumdist(\redshift)^{-1}
\end{equation}
which yields
\begin{equation}
    \label{eq:SNR_inv_relation}
    \snr(\mdeta, \mdetb, \lumdist(\redshift')) = \frac{\lumdist(\redshift)}{\lumdist(\redshift')} \snr(\mdeta, \mdetb, \lumdist(\redshift))
\end{equation}
for some arbitrary redshifts, $\redshift$ and $\redshift'$

The key here, is that we don't need to construct a three-dimensional model. 
We can interpolate in $\mdeta$ and $\mdetb$, at some reference redshift, 
    $\redshift_r$, and use the ratio of the luminosity distance
    to evaluate at an arbitrary redshift.
In doing so, we construct a function, $f(\msrca, \msrcb, \redshift)$, which draws from
    $\lalsnr(\mdeta, \mdetb, \redshift_r)$.

We define $f(\msrca, \msrcb, \redshift)$ as such:
\begin{equation}\label{eq:f_definition}
\begin{split}
f(\msrca, \msrcb, \redshift) = \\
    \snr((\redshift+1)\msrca, (\redshift+1)\msrcb, \lumdist(\redshift))
\end{split}
\end{equation}
\begin{equation}
\begin{split}
f(\msrca, \msrcb, \redshift') = \\
\snr((\redshift'+1)\msrca, (\redshift'+1)\msrcb, \lumdist(\redshift'))
\end{split}
\end{equation}

From \ref{eq:SNR_inv_relation}, we have

\begin{equation} \label{eq:f_z_transformed}
\begin{split}
    \snr((\redshift+1)\msrca, (\redshift+1)\msrcb, \lumdist(\redshift)) = \\
    \frac{\lumdist(\redshift')}{\lumdist(\redshift)} \snr((\redshift+1)\msrca, (\redshift+1)\msrcb, \lumdist(\redshift'))
\end{split}
\end{equation}

By inputting adjusted $M_{src}$ values, we can reference
    the same detector mass values at a different redshift:

\begin{equation}\label{eq:f_zref_adj}
\begin{split}
    f(\frac{\redshift + 1}{\redshift' + 1}\msrca, \frac{\redshift + 1}{\redshift' + 1}\msrcb, \redshift') =\\
    \snr((\redshift+1)\msrca, (\redshift+1)\msrcb, \lumdist(\redshift'))
\end{split}
\end{equation}

Combining \ref{eq:f_definition}, \ref{eq:f_z_transformed}, and \ref{eq:f_zref_adj},
yields:

\begin{table}
    \centering
    \begin{tabular}{|c|c|c|c|}
        \hline
        Resolution ($n$) & Grid Points ($n^2$) & sample time & 
            $Err^*$\\
        \hline
        51 & 2,601 & 2.61 sec & 0.08 \\
        101 & 10,201 & 10.9 sec & 0.037 \\
        \hline
    \end{tabular}
    \caption{Performance of the Gaussian Process model}
        Sample time is the time required (after training) to evaluate
            10,000 SNR samples, using the Gaussian Process model.
        Batches of this size reduce the memory which must
            be used to evaluate samples.
        Err is the maximum fractional error of the model
        ($max(\frac{\mathrm{model} - \snr}{\snr})$),
        tested by
            sampling the interpolation randomly.
        
    \label{tab:performance}
\end{table}

\begin{equation}\label{f_relation}
\begin{split}
    f(\msrca, \msrcb, \redshift) = \\\frac{\lumdist(\redshift')}{\lumdist(\redshift)} 
        f(\frac{\redshift+1}{\redshift'+1}\msrca, \frac{\redshift+1}{\redshift'+1}\msrcb, \redshift')
\end{split}
\end{equation}
Finally, drawing samples from a model trained at fixed $\redshift_r$ is evaluated as such,
\begin{equation}\label{f_sampler}
\begin{split}
f(\msrca, \msrcb, \redshift) = \\
\frac{\lumdist(\redshift_r)}{\lumdist(\redshift)} \snr((\redshift+1)\msrca, (\redshift+1)\msrcb, \lumdist(\redshift_r))
\end{split}
\end{equation}
We therefore train $\snr(\mdeta, \mdetb, \lumdist(\redshift_r))$ using a 
    Gaussian Process Regression model, which only requires a grid in two dimensions.

To perform Gaussian process regression more efficiently, we have implimented a 
    piecewise polynomial covariance function with compact support.
    \cite{williams2006gaussian}
These basis functions are guarenteed to be positive definite, and the
    covariance between points becomes zero as their distance increases, 
    and are given as $K_{ppD,q}(r)$.

\begin{align}
K_{ppD,0}(r) &= (1-r)^{j}_{+} \\
K_{ppD,1}(r) &= (1-r)^{j + 1}_{+} ((j + 1)r + 1) \\
K_{ppD,2}(r) &= \frac{(1-r)^{j + 2}_{+} ((j^2 + 4j +3)r^2 + (3j +6)r + 3)}{3} \\
K_{ppD,3}(r) &= (1-r)^{j + 3}_{+} 
  \left ((j^3 + 9j^2 + 23j + 15)r^3  \nonumber
   \right. \\ & \left. + (6j^2 + 36j + 45)r^2  \nonumber
  \right. \\ & \left. + (15j + 45)r +15) 
  \right )\diagup 15
\end{align}
Where $j = \lfloor\frac{D}{2}\rfloor + q + 1$, $D$ is the dimensionality of your data set.
    $q$ is chosen such that the sample function is $2q$ times differentiable.
We have chosen $q = 1$, and added a whitenoise kernel as well.
We have seen that the sample time for this function scales only with $n$ for high $n$.

\section{Properties of the Bounded Multivariate Normal Distribution in Formation Parameters}
\label{ap:FormationGaussian}
\begin{table*}[!ht]
\begin{tabular}{|c|c|c|c|c|}
\hline
& $\STFa$ & $\STBeta$ & $\STkick$ (km/s) & $\STWindOne$ \\
\hline
$\mu$ & 0.900 & 0.75 & 105.5 & 0.346 \\
\hline
$\sigma$ & 0.019 & 0.18 & 10.8 & 0.022 \\
\hline
$\rho_{\STFa}$  & 1. & -0.275 & 0.050 & 0.271 \\
$\rho_{\STBeta}$ & -0.275 & 1. & 0.830 & -0.714 \\
$\rho_{\STkick}$ & 0.050 & 0.830 & 1. & -0.624 \\
$\rho_{\STWindOne}$ & 0.271 & -0.714 & -0.624 & 1. \\
\hline
\end{tabular}
\caption{\textbf{Properties of the bounded multivariate normal distribution
    fit to models K0100-K0563};
    \label{tab:norm-properties}
    $\mu$ describes the location of the maximum likelihood for
    the multivariate normal distribution.
    $\sigma$ describes the scale of the distribution.
    $\rho$ describes the symmetric correlation matrix.
}
\end{table*}
 
Table \ref{tab:norm-properties} describes the properties of the bounded multivariate
    normal distribution used to fit the joint likelihood
    (as seen in Figure \ref{fig:formation-gaussian}).
The process of fitting a bounded multivariate normal distribution
    to a grid of likelihoods is accomplished using
    methods similar to those outlined for describing the likelihood of individual
    events (most similar to the low-latency section of \cite{nal-methods-paper}).
To expound upon this,
    the parameters of a multivariate normal distribution can be optimized
    the same way as any other parametric model,
    and we use Scipy \cite{2020SciPy-NMeth}
    for this optimization
    while using a decomposition of the multivariate normal parameters
    which lends itself well to the guarantee of positive-definite eigenvalues
    in the covariance.
When optimizing this bounded multivariate normal distribution,
    we must select a subset of our simulation models
    in order to avoid contaminating our model of the 
    subspace near our local maximum;
    this is accomplished by fitting the simulations within some range,
    $\delta \pprob$, of the maximum value of the joint likelihood.
At each iteration, this parameter is tuned by hand.
In the final iteration, $\delta \pprob = 7$, which includes
    the 27 simulations with the highest joint likelihood.

\section{Properties of the Simulated Universes}
\label{ap:tables}
We present the formation parameters, detection rates,
    and likelihoods for each model in the K series of StarTrack models,
    using the assumed cosmology.
See Section \ref{sec:models} for an interpretation of the   
    assumptions for each simulation in this series of models.
See Table \ref{tab:K_parameter_sampling} for details about how
    the formation parameter space is sampled.

\begin{widetext}
\begin{longtable}{|p{1.0cm}|p{0.8cm}|p{0.8cm}|p{1.1cm}|p{0.8cm}|p{0.8cm}|p{1.1cm}|p{1.1cm}|p{1.1cm}|p{1.2cm}|p{1.2cm}|p{1.2cm}|}
\hline
model &  $\STFa$ &  $\STBeta$ &  $\STkick$ &  $\STWindOne$ &  $\STWindTwo$ &  $\mu_{\mathrm{\bbh{}}}$ &  $\mu_{\mathrm{\bns{}}}$ &  $\mu_{\mathrm{\nsbh{}}}$ &  lnL\_rate &  lnL\_shape &  lnL\_joint \\
\hline
K0100 &  0.601 &  0.934 &  19.913 &  0.802 &  1.000 &  211.283 &  3.653 &  12.227 &  165.865 &  -220.574 &  -54.709 \\
\hline
K0101 &  0.760 &  0.295 &  240.759 &  0.882 &  1.000 &  5.419 &  1.101 &  0.092 &  110.485 &  -290.909 &  -180.424 \\
\hline
K0102 &  0.878 &  0.080 &  225.610 &  0.777 &  1.000 &  6.518 &  0.999 &  0.116 &  123.017 &  -264.868 &  -141.851 \\
\hline
K0103 &  0.254 &  0.633 &  50.807 &  0.301 &  1.000 &  138.841 &  6.213 &  5.306 &  211.830 &  -227.597 &  -15.767 \\
\hline
K0104 &  0.341 &  0.717 &  87.670 &  0.848 &  1.000 &  42.216 &  5.498 &  1.749 &  224.546 &  -267.696 &  -43.150 \\
\hline
K0105 &  0.427 &  0.379 &  110.460 &  0.927 &  1.000 &  28.240 &  5.320 &  0.946 &  209.259 &  -282.192 &  -72.934 \\
\hline
K0106 &  0.988 &  0.459 &  29.083 &  0.816 &  1.000 &  156.488 &  1.837 &  9.295 &  201.870 &  -203.949 &  -2.078 \\
\hline
\end{longtable}
 \end{widetext}

\end{document}